\documentclass[aps,preprint,prl,showpacs]{revtex4}
\usepackage{graphicx}
\begin{document}

\def \reofour {$\mathrm{(TMTSF)_{2}ReO_4}$\,}
\def \clofour {$\mathrm{(TMTSF)_{2}ClO_4}$\,}
\def \areofour {$\mathrm{ReO_4^-}$\,}
\def \aclofour {$\mathrm{ClO_4^-}$\,}
\def \pfsix {$\mathrm{(TMTSF)_{2}PF_6}$\,}
\def \asfsix {$\mathrm{(TMTSF)_{2}AsF_{6}}$}
\def \tmtsfx {$\mathrm{(TMTSF)_{2}X}$\,}

\def\tc{$T_{c}$\,}
\def\pc{$P_{c}$\,}

\def\t6as{$\mathrm{(TMTSF)_{2}AsF_{6}}$}
\def\tmx{$\mathrm{(TMTSF)_{2}ClO_{4(1-x)}ReO_{4x}}$}\,
\def\tmc{$\mathrm{(TMTSF)_{2}ClO_{4}}$\,}
\def\tms{$\mathrm{(TMTSF)_{2}AsF_{6(1-x)}SbF_{6x}}$}\,
\def\tmps{$\mathrm{(TMTTF)_{2}PF_{6}}$\,}
\def\tmttfsbf6{$\mathrm{(TMTTF)_{2}SbF_{6}}$\,}
\def\tmttfasf6{$\mathrm{(TMTTF)_{2}AsF_{6}}$\,}
\def\tmttfbf4{$\mathrm{(TMTTF)_{2}BF_{4}}$\,}
\def\tmtsfreo4{$\mathrm{(TMTSF)_{2}ReO_{4}}$\,}
\def\tq{$\mathrm{TTF-TCNQ}$\,}
\def\tsq{$\mathrm{TSeF-TCNQ}$}\,
\def\qnq{$(Qn)TCNQ_{2}$}\,
\def\R{$\mathrm{ReO_{4}^{-}}$}  
\def\C{$\mathrm{ClO_{4}^{-}}$}
\def\P{$\mathrm{PF_{6}^{-}}$}
\def\tqr{$\mathrm{TCNQ^\frac{\cdot}{}}$\,}
\def\nmpq{$\mathrm{NMP^{+}(TCNQ)^\frac{\cdot}{}}$\,}
\def\q{$\mathrm{TCNQ}$\,}
\def\nmp{$\mathrm{NMP^{+}}$\,}
\def\f{$\mathrm{TTF}\,$}
\def\tc{$T_{c}$\,}
\def\tc{$T_{c}$\,}
\def\nmq{$\mathrm{(NMP-TCNQ)}$\,}
\def\ts{$\mathrm{TSeF}$}
\def\tsm{$\mathrm{TMTSF}$\,}
\def\tst{$\mathrm{TMTTF}$\,}
\def\tmp6{$\mathrm{(TMTSF)_{2}PF_{6}}$\,}
\def\tms2x{$\mathrm{(TMTSF)_{2}X}$}
\def\as{$\mathrm{AsF_{6}}$}
\def\sb{$\mathrm{SbF_{6}}$}
\def\pf{$\mathrm{PF_{6}}$}
\def\re{$\mathrm{ReO_{4}}$}
\def\ta{$\mathrm{TaF_{6}}$}
\def\cl{$\mathrm{ClO_{4}}$}
\def\4fb{$\mathrm{BF_{4}}$}
\def\ttdm{$\mathrm{(TTDM-TTF)_{2}Au(mnt)_{2}}$}
\def\edt{$\mathrm{(EDT-TTF-CONMe_{2})_{2}AsF_{6}}$}
\def\tfx{$\mathrm{(TMTTF)_{2}X}$\,}
\def\tsx{$\mathrm{(TMTSF)_{2}X}$\,}
\def\ttftcnq{$\mathrm{TTF-TCNQ}$\,}
\def\ttf{$\mathrm{TTF}$\,}
\def\tcnq{$\mathrm{TCNQ}$\,}
\def\bedtttf{$\mathrm{BEDT-TTF}$\,}
\def\reo4{$\mathrm{ReO_{4}}$}
\def\bedtttfreo4{$\mathrm{(BEDT-TTF)_{2}ReO_{4}}$\,}
\def\et2i3{$\mathrm{(ET)_{2}I_{3}}$\,}
\def\et2x{$\mathrm{(ET)_{2}X}$\,}
\def\ket2x{$\mathrm{\kappa-(ET)_{2}X}$\,}
\def\cuncnbr{$\mathrm{Cu(N(CN)_{2})Br}$\,}
\def\ket2x{$\mathrm{\kappa-(ET)_{2}X}$\,}
\def\cuncncl{$\mathrm{Cu(N(CN)_{2})Cl}$\,}
\def\cuncs{$\mathrm{Cu(NCS)_{2}}$\,}
\def\betsfecl4{$\mathrm{(BETS)_{2}FeCl_{4}}$\,}
\def\bets{$\mathrm{BETS}$\,}
\def\hc2{$H_{c2}(T)$}
\def\et{$\mathrm{ET}$\,}
\def\tmm{$\mathrm{TM}$\,}
\def\tmtsf{$\mathrm{(TMTSF)}$\,}
\def\tmttf{$\mathrm{(TMTTF)}$\,}
\def\tm2x{$\mathrm{(TM)_{2}X}$\,}
\def\t1{${1/T_1}$\,}
\title{Linear-T scattering and pairing from antiferromagnetic fluctuations in the \tms2x organic superconductors}
\author{N.~Doiron-Leyraud, S. Ren\'e de Cotret, A.  Sedeki, C.~Bourbonnais and L.~Taillefer }
\email{ndl@physique.usherbrooke.ca}
\email{cbourbon@physique.usherbrooke.ca}
\email{ltaillef@physique.usherbrooke.ca}
\affiliation{D\'epartement de Physique and RQMP, Universit\'e de Sherbrooke, Sherbrooke, Qu\'ebec, J1K 2R1,  Canada}

\author{P. Auban-Senzier and D. J\'erome}
\email{senzier@lps.u-psud.fr}
\email{jerome@lps.u-psud.fr}
\affiliation{Laboratoire de Physique des Solides, UMR 8502 CNRS Universit\'e Paris-Sud, 91405 Orsay, France}

\author{K. Bechgaard}
\affiliation{Department of Chemistry, H.C. {\O}rsted Institute, Copenhagen, Denmark}

\date{\today}

\begin{abstract}

An exhaustive investigation of  metallic electronic  transport and superconductivity of organic superconductors \tmc and \tmp6  in the Pressure-Temperature phase diagram between $T=0$ and $20$ K and a theoretical description based  on the weak coupling renormalization group method are reported. The analysis of the data reveals a high temperature domain ($T\approx 20$ K) in which a regular $T^2$ electron-electron Umklapp scattering obeys a Kadowaki-Woods law and a low temperature regime ($T< 8$ K) where the resistivity is dominated  by a linear-in temperature component. In both compounds a correlated behavior  exists between the linear  transport and the extra nuclear spin-lattice relaxation due to antiferromagnetic fluctuations. In addition, a tight connection is clearly established between linear transport and \tc. We propose a theoretical description of the  anomalous  resistivity based on a weak coupling renormalization group determination of electron-electron scattering rate.  A linear resistivity is found  and  its origin lies  in antiferromagnetic correlations  sustained by Cooper pairing  via constructive interference. The decay of the linear resistivity  term under pressure is correlated  with  the  strength of antiferromagnetic spin correlations and $T_c$,   along with an unusual  build-up of the Fermi liquid scattering.   The results  capture the key features of  the low temperature electrical transport in the Bechgaard salts.

\end{abstract}


\maketitle
\section{Introduction}
A recent  extensive study of the transport properties has been carried on  in the most popular organic   superconductors namely the Bechgaard salts, \tmp6\cite{Jerome80} and \tmc\cite{Bechgaard81} as a function of  pressure\cite{Doiron09,Doiron10}. This previous study has focused on the electronic transport at low temperature in the limit $T\rightarrow 0$ revealing the existence of a linear temperature dependence of the resistivity \textit{at variance} with the sole $T^2$ dependence expected from the electron-electron scattering in a conventional Fermi liquid. Furthermore, the  study has established a  correlation between  the amplitude of the prefactor $A$ of the  linear temperature dependence of the resistivity observed at low temperature and the value of the superconducting critical temperature \tc.
Such a correlation between $A$ and \tc has suggested in turn a common origin for  the scattering and pairing in \tmp6\cite{Doiron09a}, both rooted in the low frequency antiferromagnetic fluctuations, as   detected by NMR experiments\cite{Wzietek93,Brown08,Bourbonnais09}. Hence, the superconducting phase of \tmp6 might be controlled by AF fluctuations with high pressure acting on the strength of correlations. 

A preliminary extension of the transport analysis up to higher temperatures ($\approx 30$ K) in \tmp6\cite{Doiron10}  has suggested that the linear law is actually  the low temperature limit of a more complex behaviour. 
 In addition, Ref\cite{Doiron09}  has pointed out a close correlation between the anomalous resistivity and the existence of an enhanced  nuclear relaxation due to antiferromagnetic fluctuations. Although transport experiments  had also been undertaken on \tmc, the ambient pressure analogue of \tmp6, leading to somewhat similar conclusions\cite{Doiron09a} a detailed analysis of the data is still missing. It is the goal of the present work.

While considered as   belonging  to the same family of organic superconductors, \tmc is different from \tmp6 in some respects. It is already known from previous studies that superconductivity (SC) in \tmc can be suppressed either by pressure\cite{Parkin85} or by a controlled amount of non magnetic defects provided by alloying or  residual disorder of non spherical \cl anions\cite{Tomic83c,Joo04}.

The proximity between SC and  SDW states is well established in the $P-T$ phase diagram of \tmp6 and a SDW phase exhibiting insulating properties can also be stabilized in \tmc under ambient pressure provided the sample is cooled fast enough to fulfill the so called quenched condition\cite{Tomic83b,Ishiguro83}.
Given the similarities and the differences between \tmc and \tmp6 regarding the  problem of anions which is relevant in \tmc  only, it is important to analyze using the  transport data, the  roles of correlations (or pressure) and anions acting as  parameters controlling the stability of SC in \tmc.

Electrical transport studies were paralleled and preceded on  the theoretical side by  a number of analysis making use of the weak coupling renormalization group (RG) approach to  the  description  of the Bechgaard salts series \cite{Duprat01,Nickel06,Bourbonnais88}. In the framework of the quasi-one-dimensional electron gas model\cite{Emery82},  these investigations  provide a rather coherent picture
 of the mechanisms of instability of the metallic state toward the onset of long-range order in  these materials. The RG calculations performed at the one-loop level showed that density-wave and Cooper pairings do not act as separate   entities in perturbation theory. They mix and interfere  at every order, a reciprocity of many-body processes  that  reproduces   the SDW-SC sequence of instabilities in  the $P-T$ phase diagram of \tms2x under pressure. The same model   proved to be also  successful    in describing the anomalous  temperature dependence of the nuclear spin relaxation rate\cite{Bourbonnais09}, $T_1^{-1}$, which stands out as a common characteristic of these materials above $T_c$\cite{Wzietek93,Brown08}. It provided   a microscopic explanation of the Curie-Weiss  enhancement  of $T_1^{-1}$\cite{Wu05}, in terms of SDW fluctuations fueled   by Cooper pairing   in  the metallic phase, linking then the size of $T_c$  to the amplitude of spin fluctuations under pressure.

The extent to which the physics of the    very same model and approach  can throw light on the origin of non Fermi liquid electron  transport above $T_c$, constitutes a clear-cut objective  for the theory.  A key ingredient for resistivity   is the electron-electron scattering rate on the Fermi surface, a quantity that  can be extracted  from  the calculation of the  one-particle self-energy. It requires an extension of   the  RG method  up to  the  two-loop   level, a program that has been carried out recently.\cite{Sedeki10}    The results  will be used in a calculation of resistivity and  their applicability  to   electrical  transport experiments for (TMTSF)$_2$PF$_6$ and (TMTSF)$_2$ClO$_4$ attested.

\section{Experimental}
 In the present paper  we report on measurements of the  electrical resistivity in \tmc  and \tmp6 mostly  along the $a$-axis i.e, along the chains of organic molecules, as a function of pressure and temperature. 
 Single crystals  were grown by the usual method of electrocrystallisation~\cite{Bechgaard79}. 
 Typical sample dimensions are 1.5 x 0.2 x 0.05 mm$^3$ which are the length, width and thickness along the $a$, $b$, and $c$ crystallographic axes, respectively. Electrical contacts were made with evaporated gold pads (typical resistance between 1 and 10 $\Omega$) to which 17 $\mu$m gold wires were glued with silver paint. The current was applied along the $a$-axis.The magnetic field was aligned with  the $c^{\star}$-axis.  The electrical resistivity was measured with a resistance bridge using a standard four-terminal AC technique. Low excitation currents of typically 30 $\mu$A were applied in order to eliminate heating effects caused by the contact resistances. 
  The samples used have typical values of $a$-axis conductivity near $500 (\Omega$ cm)$^{-1}$ and $400 (\Omega$ cm)$^{-1}$ for \tmc and \tmp6 respectively. 
 A non-magnetic piston-cylinder pressure cell was employed\cite{Walker99}, with Daphne oil as pressure transmitting medium. The pressure at room temperature and 4.2 K was measured using the change in resistance and superconducting $T_c$ of a Sn sample, respectively. Only the values recorded at 4.2 K are quoted here. The low temperature down to 0.1 K was provided by a demagnetisation   fridge.
From room temperature down to 77 K the cooling rate was kept below
1 K/min to ensure gradual freezing of the pressure medium and an optimal level of
pressure homogeneity, and to avoid the appearance of cracks in our samples.  All the data reported here are on samples that
showed no sign of cracks, \textit{i.e}., their resistance at room temperature always recovered their
initial value prior to each cooling cycle. No cracks were detected during pressurization
either, \textit{i.e}., the resistance at room temperature evolved smoothly with the  applied pressure.
Below 77 K, the cooling rate was kept below 5 K/hour to ensure adequate thermal
equilibrium between the samples and the temperature sensors placed outside the cell as slow cooling is vital to optimize anion ordering  which
occurs at 25 K at low pressure in \tmc. 

\section{Results}
\subsection{SC phase diagram}
The superconducting transition
temperature \tc  was determined using the onset temperature according to the measured temperature dependence of the resistivity down to 0.1 K.
\begin{figure}[ht]
\centerline{\includegraphics[width=0.55\hsize]{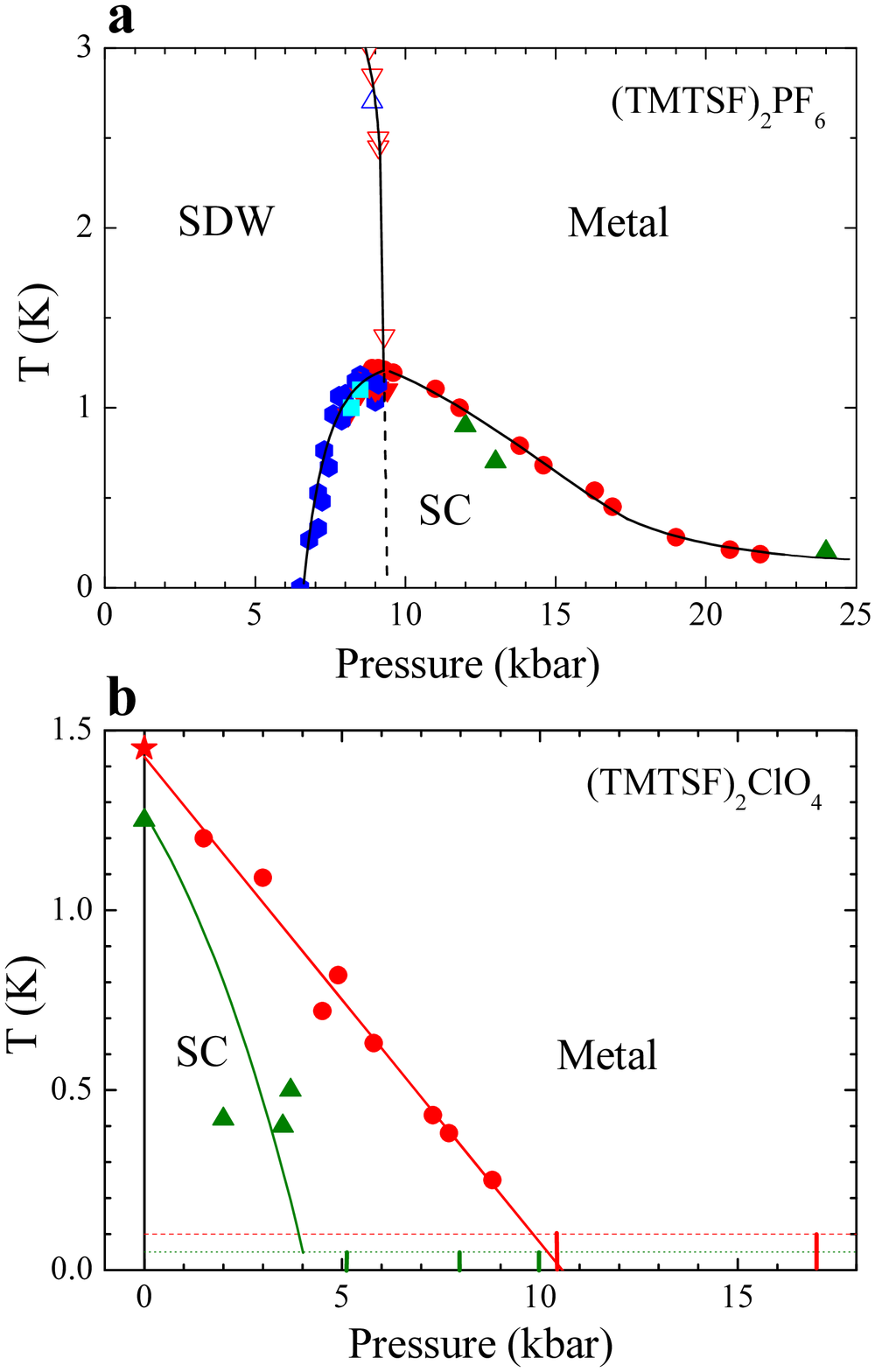}}
\caption{(a) ($P-T$) phase diagram of \tmp6 deduced from resistivity measurements. The data points below $9.4$ kbar (the coexistence regime) are deduced from measurements along the three crystallographic axes: down triangles for $\rho_a$, squares for $\rho_b$,hexagons for $\rho_c$, empty symbols for $T_{SDW}$ and full symbols for $T_{SC}$\cite{Vuletic02,Kang10}. Above the critical pressure ($9.4$ kbar), only longitudinal resistivity data are plotted: (red) circles from this work and (green) triangles from Ref.\cite{Ribault80}. (b) Pressure dependence of the superconducting transition of \tmc deduced from longitudinal resistivity measurements: (red) circles from this work (the star at 1 bar is derived from a $\rho_c$ measurement\cite{Yonezawa08}) and (green) triangles from Ref.\cite{Parkin85}. Dashed-dotted horizontal lines (red or green) indicate the lowest reached temperature without superconductivity for both studies. The (red) continuous line is a linear fit of the data including the point at 1 bar.  
\label{Fig1_DP3K}}
\end{figure}
Using such a determination for the onset of SC in transport data, the $P-T$ superconducting phase diagram has been obtained as displayed on Fig.~\ref{Fig1_DP3K}-a  for \tmp6 and Fig.~\ref{Fig1_DP3K}-b for \tmc . Note that the definition of \tc used in the present article, as the onset of the resistive transition, is different from that used in Refs.~\cite{Doiron09} and \cite{Doiron09a}, where \tc was defined as the midpoint of the resistive transition (i.e. mid-height of drop; see figure S9 in Ref.~\cite{Doiron09a}). As a result, the present \tc values are slightly higher. Also, while for \tmc  \tc(midpoint) $ \rightarrow 0$ at P $\approx 8$ kbar in Ref.\cite{Doiron09a}, and for \tmp6 \tc(midpoint) $ \rightarrow 0$ at P $\approx 22$ kbar in Refs.~\cite{Doiron09} and \cite{Doiron09a}, here we find \tc(onset) $ \rightarrow 0$ at P $\approx 10$ kbar in \tmc (Fig.~\ref{Fig1_DP3K}-b) and at P $ > 24$ kbar in \tmp6 (Fig.~\ref{Fig1_DP3K}-a). As far as \tmp6 is concerned, the coexistence region between SDW and SC is well documented and  a quantum critical point for the suppression of antiferromagnetic ordering would be located at $9.4$ kbar with the present pressure scale~\cite{Vuletic02,Kang10}. For \tmc at ambient pressure, we took  the value of \tc obtained from  a good quality run performed on a $\rho_c$ sample and cooled down to low temperature without any cracks~\cite{Yonezawa08}. All data points plotted as red circles on Fig.1 are obtained in the present work and are deduced from successive runs performed on the same sample of either \tmp6 or \tmc. 
\begin{figure}[ht]
\centerline{\includegraphics[width=12.0cm]{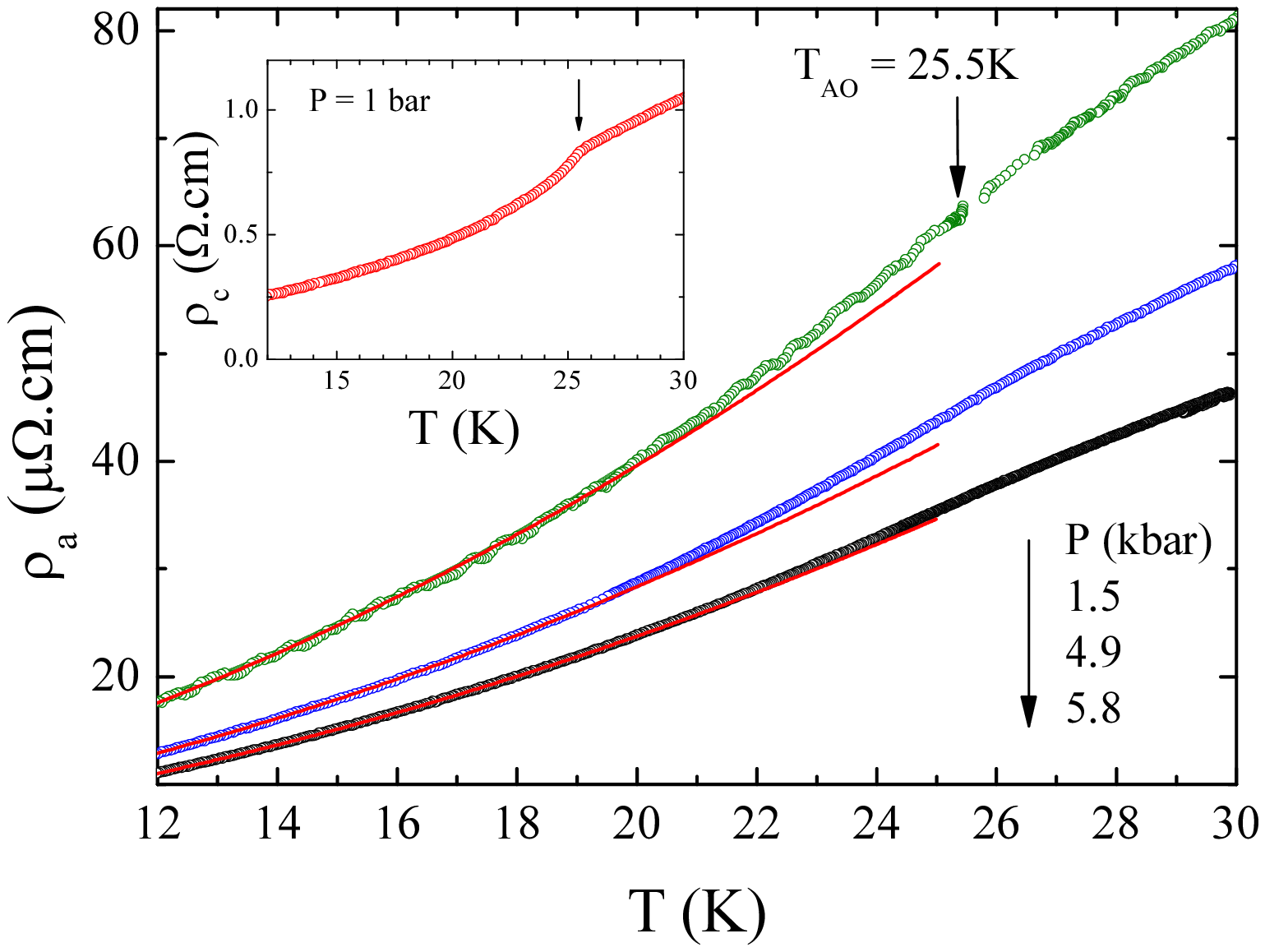}}
\caption{Temperature dependence of the longitudinal resistivity of \tmc in the vicinity of the anion ordering, $T_{AO}$, for different applied pressures; insert: Temperature dependence of the transverse resistivity of \tmc at ambient pressure in the same temperature range\cite{Auban10}.
\label{Fig2_RTAO}}
\end{figure}
Figure.~\ref{Fig1_DP3K} deserves several comments as far as  \tmc is concerned.  The most significant one is the pressure dependence of \tc which deviates strongly  from the previously published results~\cite{Parkin85}. As one cannot argue for real differences in the sample quality, all samples coming from the same chemistry lab, the difference between the two sets of data has to be found in  a more intrinsic reason namely,  the  possible disorder introduced by  \cl~anions on cooling.

There have been several reports related to the ordering of the anions in \tmc~\cite{Pouget83a,Pouget83,Pesty85a}. First, fast cooling (quenching) of the sample is known to preclude the ordering of the anions. Instead of a SC ground state, it is a SDW insulating state which becomes the stable ground state below 5 K~\cite{Tomic83b,Kagoshima83}. Second, the signature of the anion ordering on transport amounts to a drop of the resistivity at the ordering temperature $T_{AO}$, which also coincides with the onset  of the superstructure observed in X-ray diffuse scattering experiments~\cite{Pouget83a}. Furthermore, quench experiments suggest that the dynamics of anion orientation is rather slow at low temperature although this feature has yet to be  studied more quantitatively under pressure~\cite{Tomic83b,Pouget90}.

A  clear signature for the effect of anion ordering on the resistivity is provided by the transport under ambient pressure for $\rho_c$\cite{Auban10}, see insert of Fig.~\ref{Fig2_RTAO}, and under $P=1.5$ kbar for $ \rho_a$ (see Fig.~\ref{Fig2_RTAO}). 

The ordering of the anions decreases the amount of static disorder and in turn the strength of  elastic scattering of the carriers, as it can be seen by a drop  of the  resistivity below $T_{AO}$; this amounts to about $\Delta\rho \approx$ 100~{\rm m$\Omega$ cm} and   2-3 {\rm m$\Omega$ cm}  for  transverse and longitudinal transports respectively, \textit{see} Fig.~\ref{Fig2_RTAO}. 

In the present  pressure study,  special attention has been paid to the cooling conditions in order to guarantee the best possible anion ordering. As seen in Fig.~\ref{Fig2_RTAO}, a broad shoulder, although smaller in amplitude than the ambient pressure one, is still observed up to 5.8 kbar at a temperature of about 25.5~K, which can be attributed to the signature of an anion ordering still present under pressure. Our results do show that $T_{AO}$ persists above 1.5 kbar and then becomes hardly affected by pressure. This feature corroborates previous   pressure experiments showing that above the  low pressure regime where the initial pressure dependence of  $T_{AO}$ is large, the pressure coefficient becomes much smaller\cite{Murata85,Creuzet85,Guo98}.
An opposite  conclusion  namely, the  anion ordering is suppressed under pressure, had been claimed according to  a study of the angular magnetoresistance\cite{Kang93} . 

However, it looks as if  long-range ordering becomes less  perfect at low temperature under pressure  since the amplitude of the drop of the resistivity  is seen to decrease under pressure\cite{Guo98}.  Consequently, the anion ordering possibly spreads over a broader temperature regime below $T_{AO}$. We  suggest that this can be a result of anion dynamics slowing down under  pressure~\cite{Guo98}. 

Hence, we propose that the  $P-T$ phase diagram of \tmc shown  in this work is the one relevant for samples exhibiting the highest possible degree of anion ordering namely, the same  cooling rate  in the vicinity of the anion ordering temperature as used in ref\cite{Guo98}. The difference between the present phase diagram and the one inferred from the data of 1985 can be attributed to  cooling conditions  not being slow enough in the  early experiments. Actually, the effect of the cooling rate on \tc has been studied at ambient pressure in some details. Increasing the cooling rate prevents a good ordering of the \cl \,anions and in turn depresses \tc.  Above a rate of 15~K/mn, it is the SDW phase that  becomes the stable ground state\cite{Tomic83b,Garoche82,Matsunaga99}.

We see in Fig.~\ref{Fig1_DP3K} that \tc for \tmc varies linearly with pressure and its extrapolation hits the pressure axis around 10~kbar. No \tc can be detected at $10.4$ and $17$ kbar above 0.1~K.  Such a  behavior can be ascribed to the pair breaking effect of residual anion disorder in \tmc under pressure\cite{Joo05}. In contrast, in \tmp6 a \tc of 0.2~K is still obtained under 24 kbar \cite{Ribault80,Schulz81},  \textit{i.e},  14-15~kbar beyond the critical pressure for the stabilisation of SC. \tmp6 is a compound where the disorder of the anions does not come into play and therefore  pressure is the only control parameter for SC as long as the samples do not suffer from chemical defects or other kinds of defects.

\subsection{Transport}
\subsubsection{\tmp6\label{TrPF6}}
\begin{figure}[ht]
\centerline{\includegraphics[width=0.7\hsize]{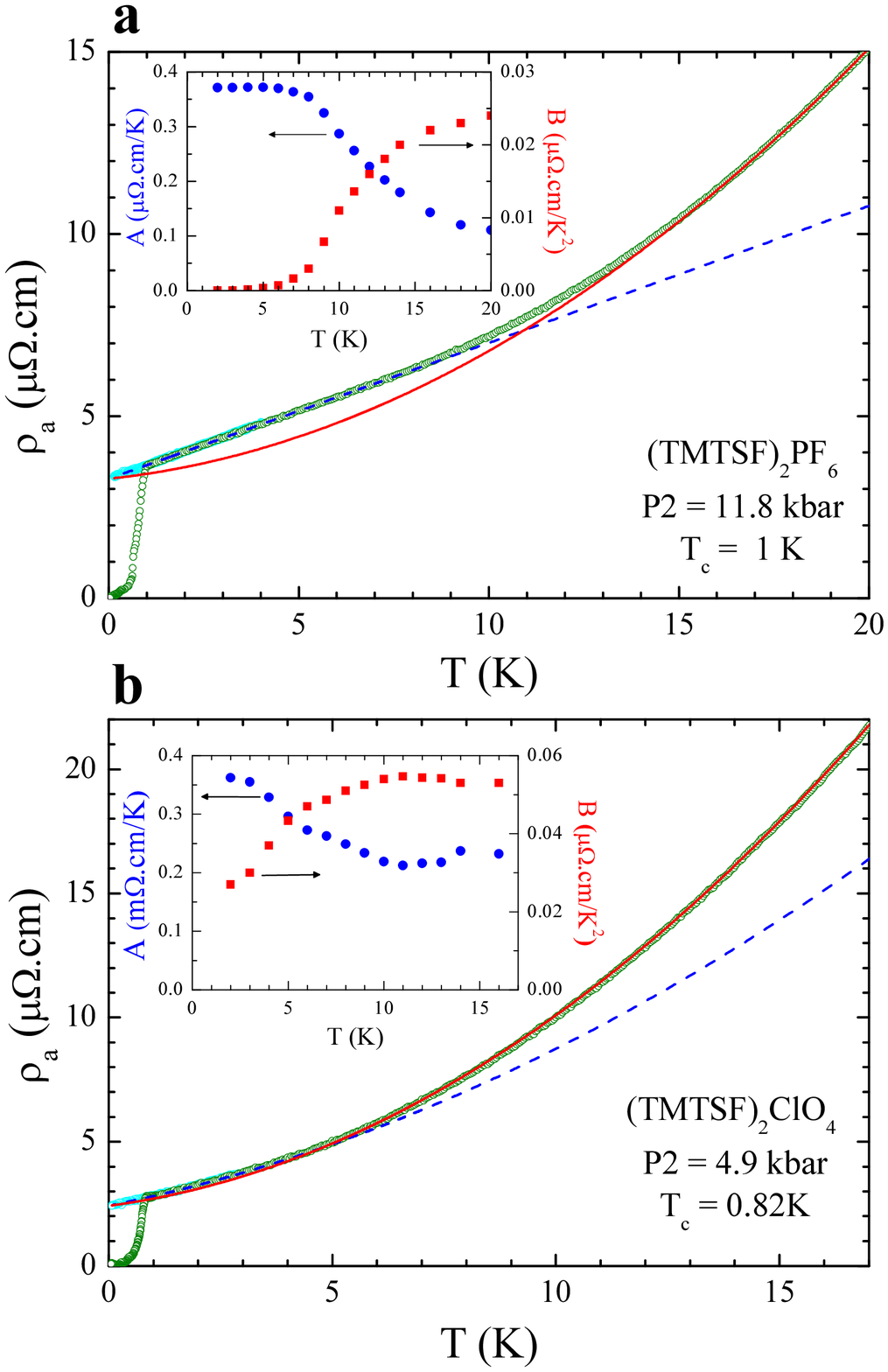}}
\caption{Temperature dependence of the longitudinal resistivity of \tmp6 at $P$ = $11.8$ kbar below $20$ K (a), and \tmc at $P$ = $4.9$ kbar below $17$ K (b), at zero field and under $H$ = $0.05$ T along $c^*$ in order to suppress SC. The second order polynomial fit, $\rho(T) = \rho_0 + A(T)T + B(T)T^2$, according to the sliding fit procedure described in the text is shown for the  $T$ intervals $(2-6)$ K and  $(18-22)$ K or $(13-17)$ K in blue and red respectively. The top inserts provide the  temperature dependence of the  $A$ and $B$ coefficients. 
\label{Fig3_RvsTABvsT}}
\end{figure}

 Figure \ref{Fig3_RvsTABvsT}-a,  displays a typical temperature dependence  up to 20 K for the longitudinal resistivity of \tmp6 under $11.8$ kbar, a pressure which is  close to the critical pressure \pc. Data are shown at zero field and under $H=0.05$ T along $c^*$ in order to suppress SC without magnetoresistance.
Below \pc =$9.4$ kbar, \tmp6 still exhibits SC features but they arise in the coexistence regime below the onset of a SDW state\cite{Kang10} which is  not relevant for the present study. 

The important feature emerging from the resistivity data in Fig.~\ref{Fig3_RvsTABvsT}-a is the  linear temperature dependence below 8 K, becoming  quadratic at higher temperatures. This quadratic contribution is absent at low temperature while the linear one becomes weaker at high temperature.
Although the  linear behaviour is the dominant feature of the resistivity at low temperature, a small saturation becomes observable   below 2 K at pressures much higher than \pc.

Taking into account  these two contributions to the resistivity we have analyzed the transport in \tmp6 fitting the experimental data in the normal state (both at zero field and under a small magnetic field) by a second order polynomial form such as $\rho(T) = \rho_0 +  A(T)T + B(T)T^2$. Here, both $A$ and $B$ prefactors can be temperature dependent, while the value of $\rho_0$ depends on pressure only. The residual resistivity$ \rho_0$ is first extracted from the fit between 0.1 and 4 K. For the determination of $A$ and $B$ a fit of the experimental data is performed at the temperature $T$ over a temperature window of $4$K centered on $T$, keeping the same value for $\rho_0$. 
The present analysis is restricted to the temperature domain below 20 K because  measurements of the transverse transport in the same materials\cite{Auban10} have shown that a cross-over from coherent  to incoherent transverse transport along $c^{\star}$ is  occurring at higher temperatures. This  may in turn affect the temperature dependence of the longitudinal resistivity since a logarithmic factor should be added to the quadratic law in the two dimensional regime when neighboring $(a, b)$ planes are decoupled\cite{Gorkov98,Hodges71,Auban10}.

The result of this analysis at 11.8 kbar is displayed on the insert of Fig.~\ref{Fig3_RvsTABvsT}-a   
namely, a coexistence of the $A$ and $B$ contributions to the transport over the whole temperature domain with a predominance for the linear contribution over the quadratic one at low temperature and the reverse at high temperature.
\begin{figure}[ht]
\centerline{\includegraphics[width=0.65\hsize]{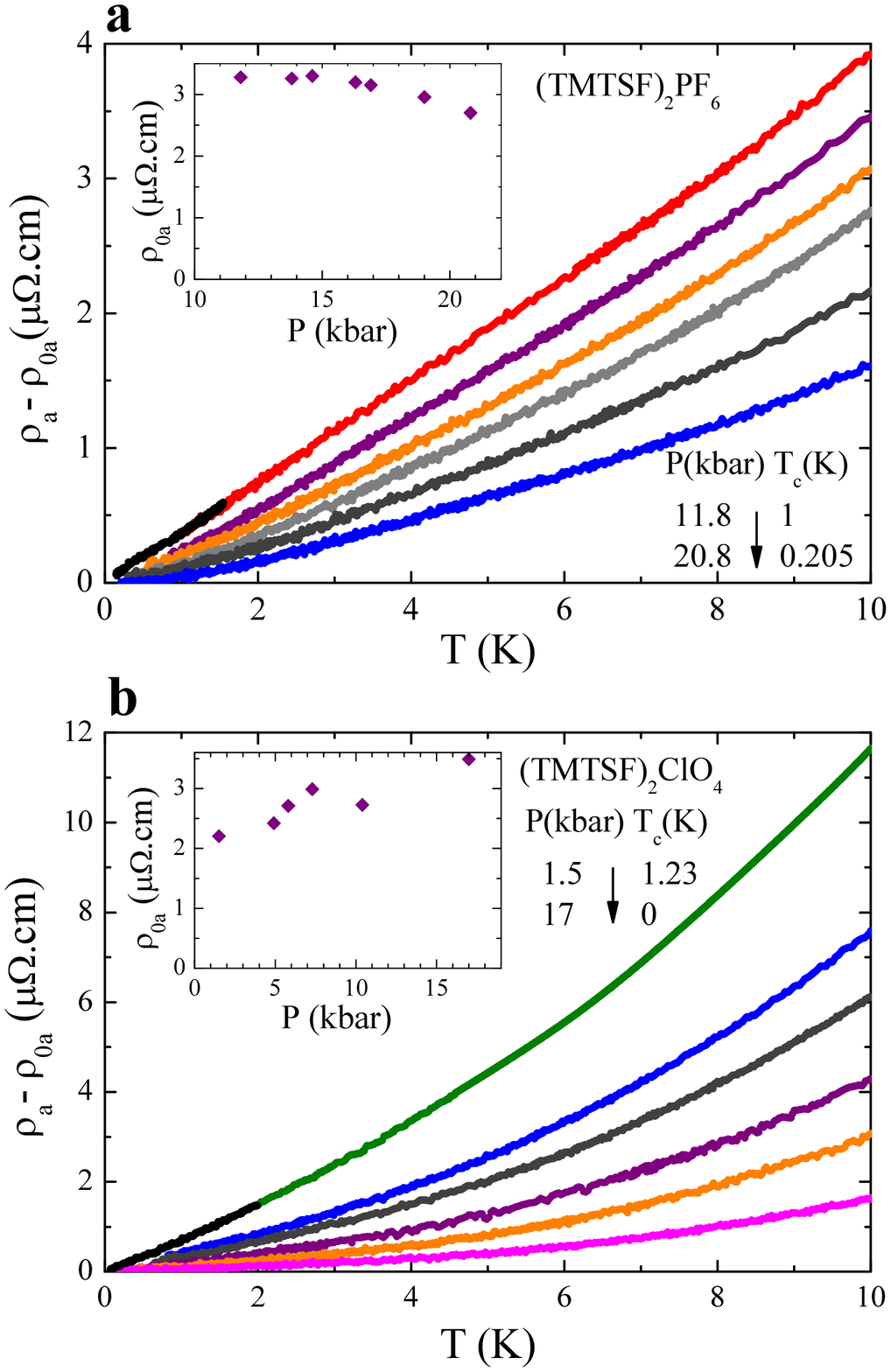}}
\caption{(color online) Inelastic contribution $\rho_a - \rho_{0a}$ of the normal state resistivity at different pressures for \tmp6 (a) and \tmc (b). Data under a small magnetic field (black) are shown for the lowest pressure only.  Inserts show the pressure dependence of the residual resistivity, $\rho_{0a}$, deduced from the low temperature fit.
\label{Fig4_DRavsTR0vsP}}
\end{figure}

Similar analyses have subsequently been conducted on transport data performed on the same sample under seven  pressures (11.8, 13.8, 14.6, 16.3, 16.9, 19 and  20.8 kbar). The inelastic resistivity data are displayed on Fig.\ref{Fig4_DRavsTR0vsP}-a and the results for  for $A(T)$ and $B(T)$  shown on Fig.~\ref{Fig5_ABT}-a after the determination of the residual resistivity under every pressure.

The pressure dependence of the residual resistivity displayed in the insert of Fig.~\ref{Fig4_DRavsTR0vsP}-a shows a smooth and weak \textit{decrease}. The smoothness of  the  pressure dependence of $\rho_0$  is an indication for the reliability of the data  of the  seven successive pressure  runs. The pressure coefficient ($-2.5\%$/kbar) can be ascribed to the expected decrease of the  effective mass under pressure \cite{Welber78,Jerome82}. 
The presence of a small saturation below 2 K manifests as a large increase of $B$ below 5 K while at the same time $A$ is slightly decreasing. This effect is relatively more pronounced for the highest pressures.
\subsubsection{\tmc\label{TrClO4}}
\begin{figure}[ht]
\centerline{\includegraphics[width=0.8\hsize]{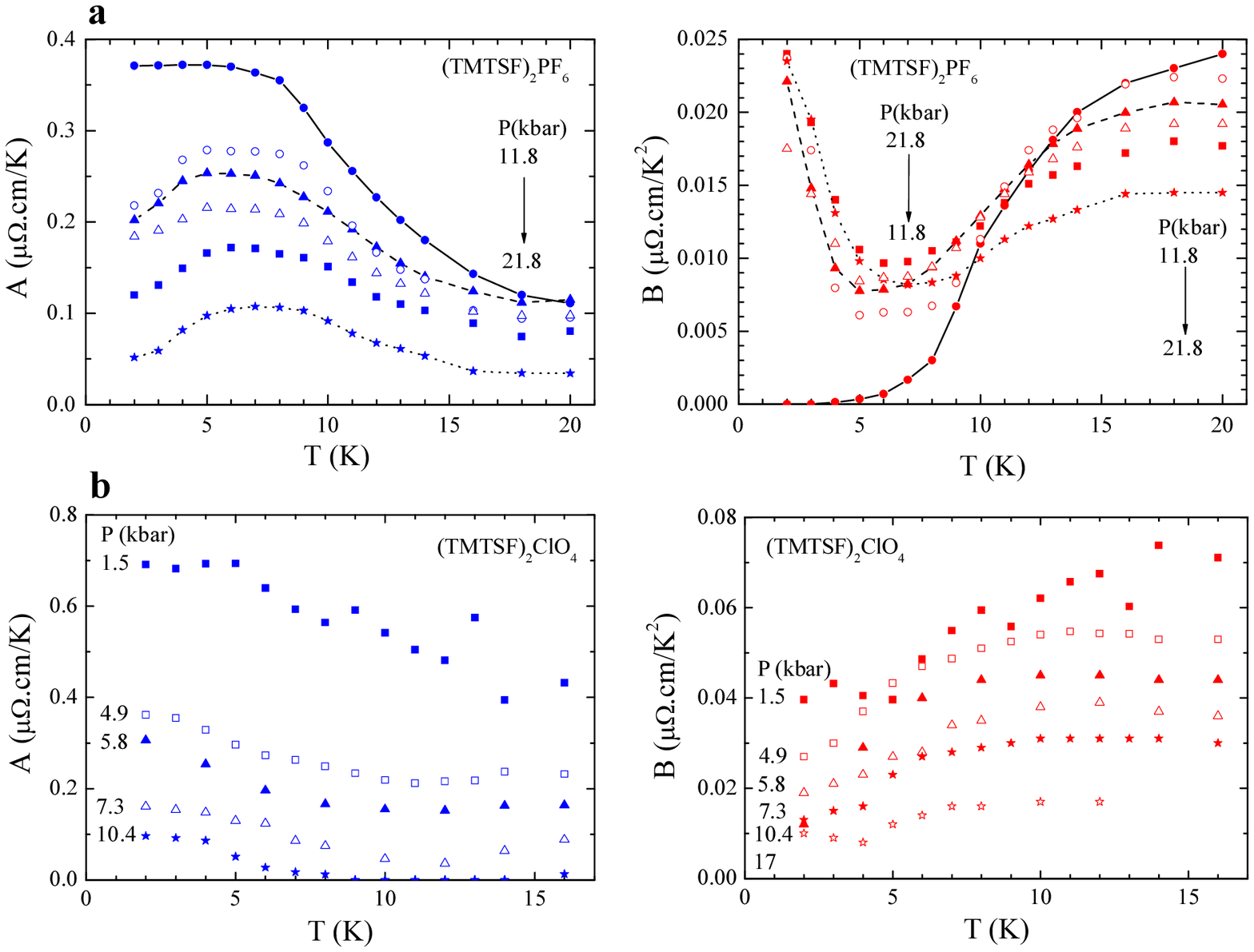}}
\caption{ Temperature dependence of $A$ (left) and $B$ (right) coefficients of the polynomial fit, $\rho(T) = \rho_0 +  AT + BT^2$  at different pressures for \tmp6 (a) and \tmc (b).
Each temperature point corresponds to the center of a $4$K window used for the fit. As far as \tmp6 is concerned, this figure shows that  under 11.8 kbar $B$ is zero  and $A$ constant below 4 K within the accuracy of the measurements and data processing.  However, a small low temperature quadratic term arises  at larger pressures, (see Fig.\ref{Fig7_BChiP}).
\label{Fig5_ABT}}
\end{figure}

A similar investigation has been performed on \tmc, although restricted to the pressure regime 1 bar-17 kbar since superconductivity is already stable at ambient pressure in this compound. A typical temperature dependence for $\rho_a$ is shown on Fig.~\ref{Fig3_RvsTABvsT}-b at the pressure of $P= 4.9$ kbar.
For the analysis of the \tmc data, using the same procedure as for \tmp6, we restrict ourselves to the temperature domain between 0.1 and 16 K since the actual temperature  dependence above 16 K is affected by the extrinsic influence of anion ordering occurring around 25~K. Unlike \tmp6, the difference between low and  high temperature regimes for \tmc is not as pronounced. A significant quadratic contribution remains at low temperature besides the dominant linear one and no additional saturation could be detected at very low temperature. Effectively, the data  for \tmc at $P= 4.9$ kbar on Fig.~\ref{Fig3_RvsTABvsT}-b  reveal a resistivity  which retains a finite temperature dependence approaching 0 K, but follows a quadratic dependence above 12 K or so. 
The analysis of the resistivity at $P= 4.9$ kbar according to the sliding fit procedure is shown in the insert of Fig.~\ref{Fig3_RvsTABvsT}-b. The results of the analysis of five consecutive pressure runs (1.5, 4.9, 5.8, 7.3 and 10.4 kbar) over ten performed on the same sample are shown on Fig.~\ref{Fig4_DRavsTR0vsP}-b for  the inelastic contribution and on Fig.~\ref{Fig5_ABT}-b for the prefactors $A$ and $B$.
The smoothness of the variation of the residual resistivity \textit{increasing} under pressure is also an indication for the good quality of the data. We can ascribe the slight increase of $\rho_0$ under pressure, instead of a decrease for \tmp6, to the anion ordering becoming less complete at $T_{AO}$  under pressure and spreading all the way down to 0 K  on account of the pressure-induced slowing down of the anion  dynamics discussed above.
\section{Discussion}
\subsection{Transport} 
\subsubsection{The $A$ contribution\label{AContr}}
\begin{figure}[ht]
\centerline{\includegraphics[width=0.7\hsize]{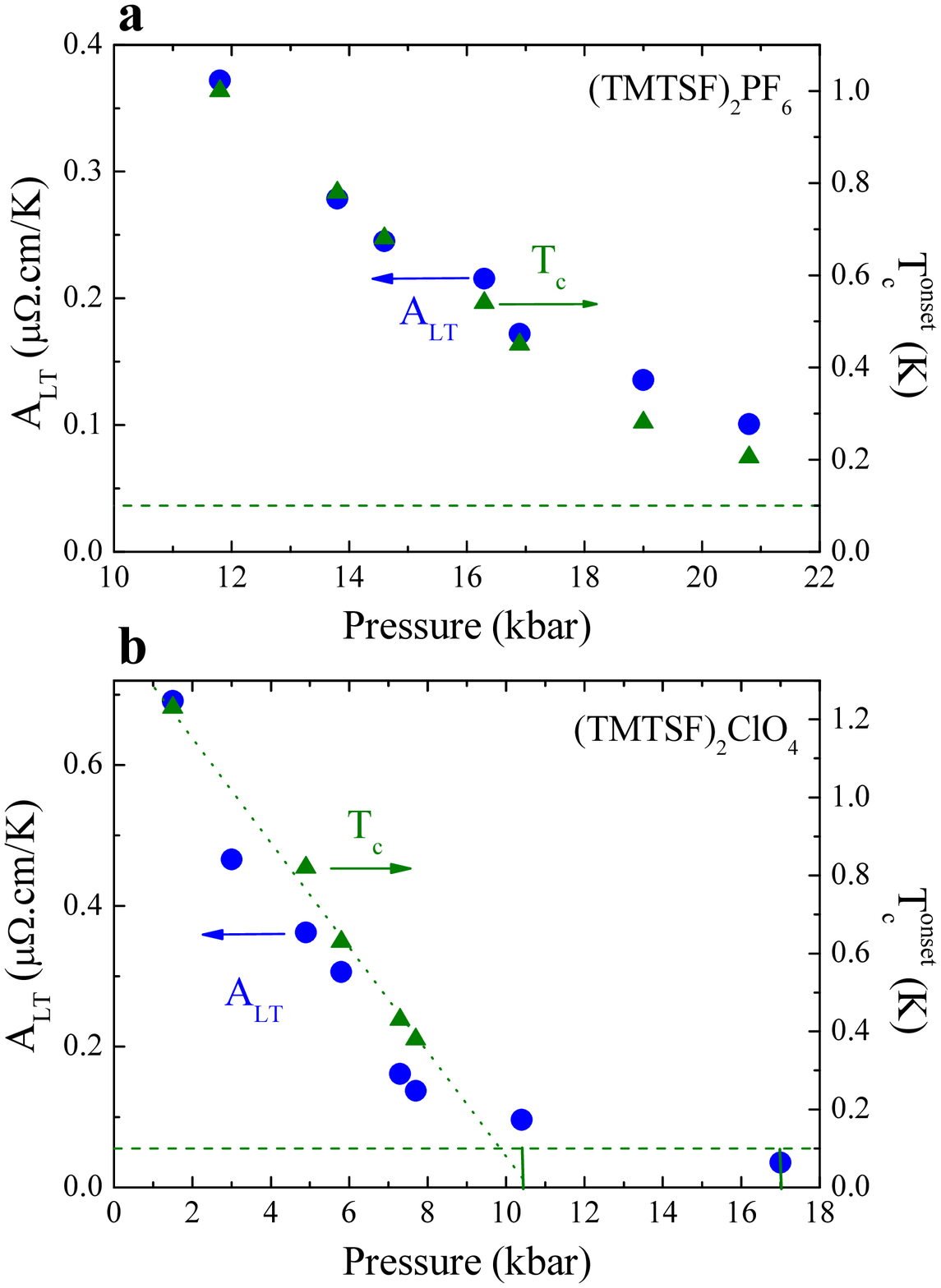}}
\caption{Pressure dependence of \tc onset and $A_{LT}$ coefficient for \tmp6 (a) and \tmc (b). $A_{LT}$ is obtained from the second order polynomial fit described in the text in the temperature window corresponding to its maximum value for \tmp6 and in the $0.1-4$ K window for \tmc. The horizontal dashed lines indicate our lowest reached temperature, $0.1$K. For \tmc, the vertical lines below this temperature indicate the pressure points at $10.4$ and $17$kbar without superconductivity and the dotted line is a linear fit of \tc onset data points.
\label{Fig6_ATcvsP}}
\end{figure}

 Figure.\ref{Fig6_ATcvsP} shows that a correlation between $A$ and \tc can be established in both compounds using $A_{LT}$,the maximum value of $A$ determined at low temperature and \tc given by the onset of SC. For \tmp6, the maximum value of $A$ is reached at a temperature which is slightly increasing with pressure (from 4 K at 11.8 kbar up to 7 K at 20.8 kbar) while for \tmc, it is always reached at the lowest temperature (0.1-4 K window). As far as \tmp6 is concerned, the pressure dependence of both quantities are nearly parallel. SC is observed up to the maximum accessible pressure (20.8 kbar) and $A_{LT}$ remains finite at such a pressure. 
The relation between $A_{LT}$ and \tc under pressure is different in \tmc. Both quantities are quite parallel in the low pressure limit, but a finite value of  $A_{LT}$  persists even in the absence of  SC  under pressure  as shown by the pressure runs at 10.4 and 17 kbar where no SC can be observed above 0.1 K \textit{at variance} with \tmp6. 
Such  differences  between  pressure dependences  in these two organic superconductors can again be explained by the particular role played by the anions in case of \tmc since  the  frozen anion disorder is affecting  the stability of the superconducting phase but not the inelastic contribution to the transport. This hypothesis is supported by the investigation of SC in solid solutions \tmx~when non magnetic point defects are introduced by alloying in a controlled way\cite{Joo04}. It can be ascribed to the pair breaking effect of non magnetic defects in a superconductor exhibiting line nodes in the gap\cite{Joo05}.

\subsubsection{The $B$ contribution}
\label{The $B$ contribution}
Unlike $A$,  no direct correlation can be established between $B$ and \tc.
The quadratic law is a well established behaviour in \tmtsfx  below $50$ K or so\cite{Cooper86,Korin88}. This law was noticed in the early studies of \asfsix under ambient pressure \cite{Tomic91} and is still valid under $20.8$ kbar in \tmp6. The same is true for \tmc even under 10.4 and 17 kbar when no SC can be observed above 0.1 K.
\begin{figure}[ht]
\centerline{\includegraphics[width=0.7\hsize]{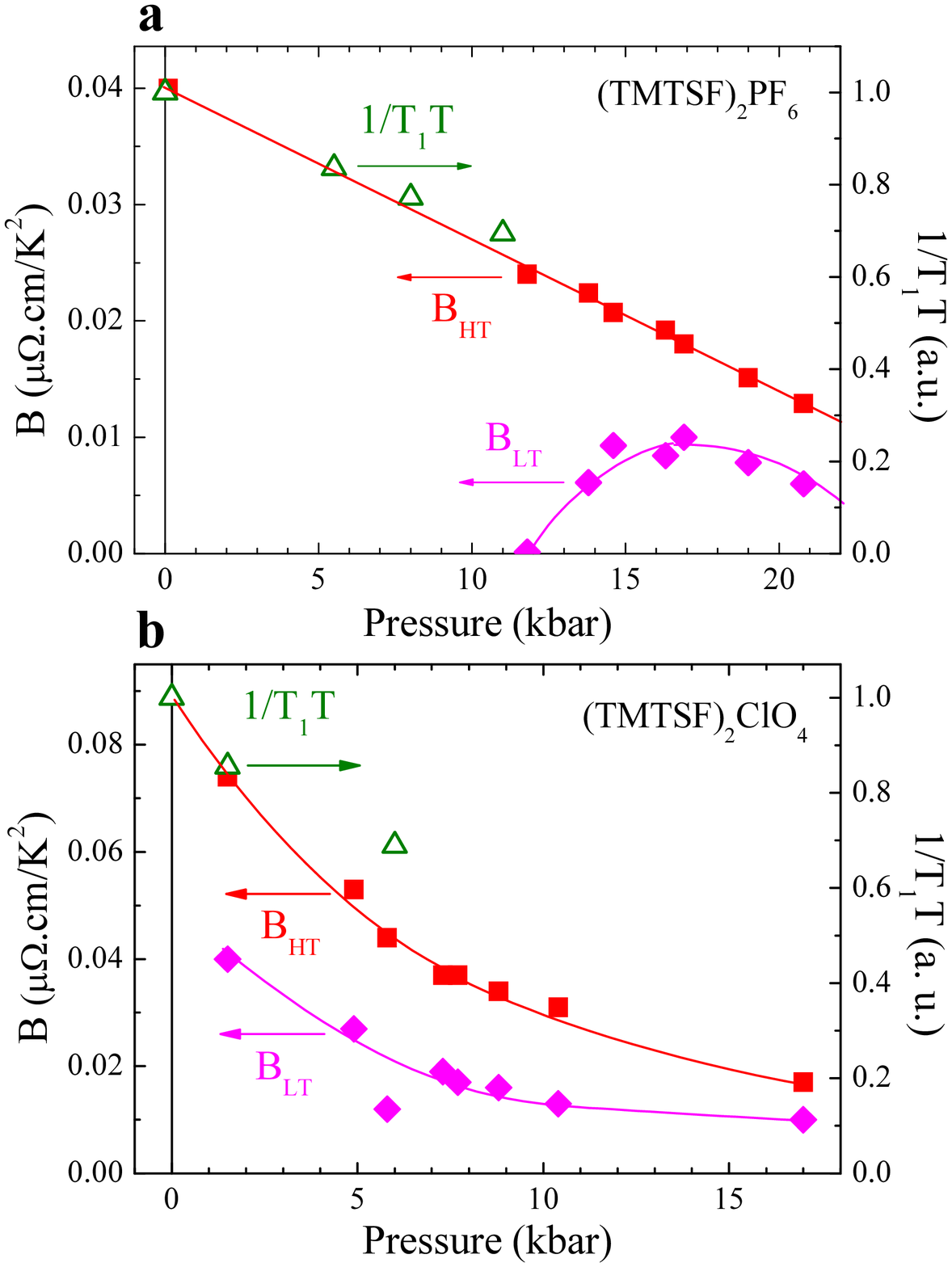}}
\caption{Pressure dependence of the $B_{HT}$ coefficient, obtained from the high temperature ($T=20$ K) polynomial fit, of the spin susceptibility measured under pressure \textit{via} NMR relaxation experiments\cite{Bourbonnais88} ($\chi^2(q=0) \propto 1/T_{1}T$) and of the $B_{LT}$ coefficient, determined in the same temperature window as $A_{LT}$ coefficient, for \tmp6 (a). The ambient pressure point for $B_{HT}$ is deduced from a-axis resistivity data in Ref.\cite{Tomic91}. Similar plot for \tmc (b) using the NMR data from Ref.\cite{Creuzet85a}. The relation $B\propto \chi^{2}$ is indicative of the Kadowaki-Woods relation encountered in strongly correlated metals\cite{Kadowaki86}.
\label{Fig7_BChiP}}
\end{figure}
Furthermore, for both compounds an experimental correlation can be established between $B$ and the electronic spin susceptibility.  As shown in Fig.~\ref{Fig7_BChiP}-a for \tmp6, the prefactor  $B_{HT}$ namely, $B$  of the polynomial law  taken at high temperature and  the square of the electronic spin susceptibility given by the NMR relaxation rate data at high temperature ($\chi^2(q=0) \propto 1/T_{1}T$)\cite{Wzietek93} follow the same pressure dependence. This behaviour is  actually reminiscent  of  the Kadowaki-Woods law observed in various strongly correlated metals\cite{Kadowaki86,Powell10}. Similar to $A$, the $B$ contribution is not affected by impurities provided by  alloying or by the frozen anion disorder\cite{Auban10}.  $B$ is controlled by pressure only. A similar comparison between $B$ and the spin susceptibility under pressure is shown in \tmc on  fig.\ref{Fig7_BChiP}-b. The quality of the agreement between NMR and transport data is not as good for \tmc.  This could be due to a large uncertainty in the knowledge of pressure at low temperature in this early NMR experiment\cite{Creuzet85a} using  argon as the pressure medium.

In addition, data of \tmp6 on Fig.~\ref{Fig5_ABT} reveal an interesting behaviour for $B_{LT}$ in the low temperature domain. At 5 K, $B_{LT}$ starting from a zero value under 11.8 kbar is increasing sharply at higher pressures and then levels off as displayed on Fig.~\ref{Fig7_BChiP}-a. It  is  actually the reason for the observation of a  linear  resistance in \tmp6 when the pressure is close to the critical pressure. We may also notice that quite a similar behaviour has derived from the measurement of $\rho_c$ at the same pressure although on different \tmp6 samples\cite{Auban10}. However, such a crossing in the temperature dependence of the $B$ coefficients seen on Fig.\ref{Fig5_ABT} is not observed in the \tmc compound where $B$ never reaches a zero value at low temperature even in the vicinity of ambient pressure. Fig.~\ref{Fig7_BChiP}-b shows a parallel evolution with pressure for $B_{LT}$ and $B_{HT}$. We may suggest several reasons for the  difference between compounds. First, the quadratic term is stronger in \tmc than in \tmp6 possibly due to a transverse coupling $t_{\perp}$ being larger   in the latter compound under pressure with $B \propto 1/t_{\perp}$ for a Q1D Fermi surface\cite{Gorkov98}. Second, the temperature dependence of $\rho$ may be spoiled in \tmc by the extended  anion ordering occurring on cooling. 
\subsection{Interplay between transport and  magnetism}
\begin{figure}[ht]
\centerline{\includegraphics[width=0.7\hsize]{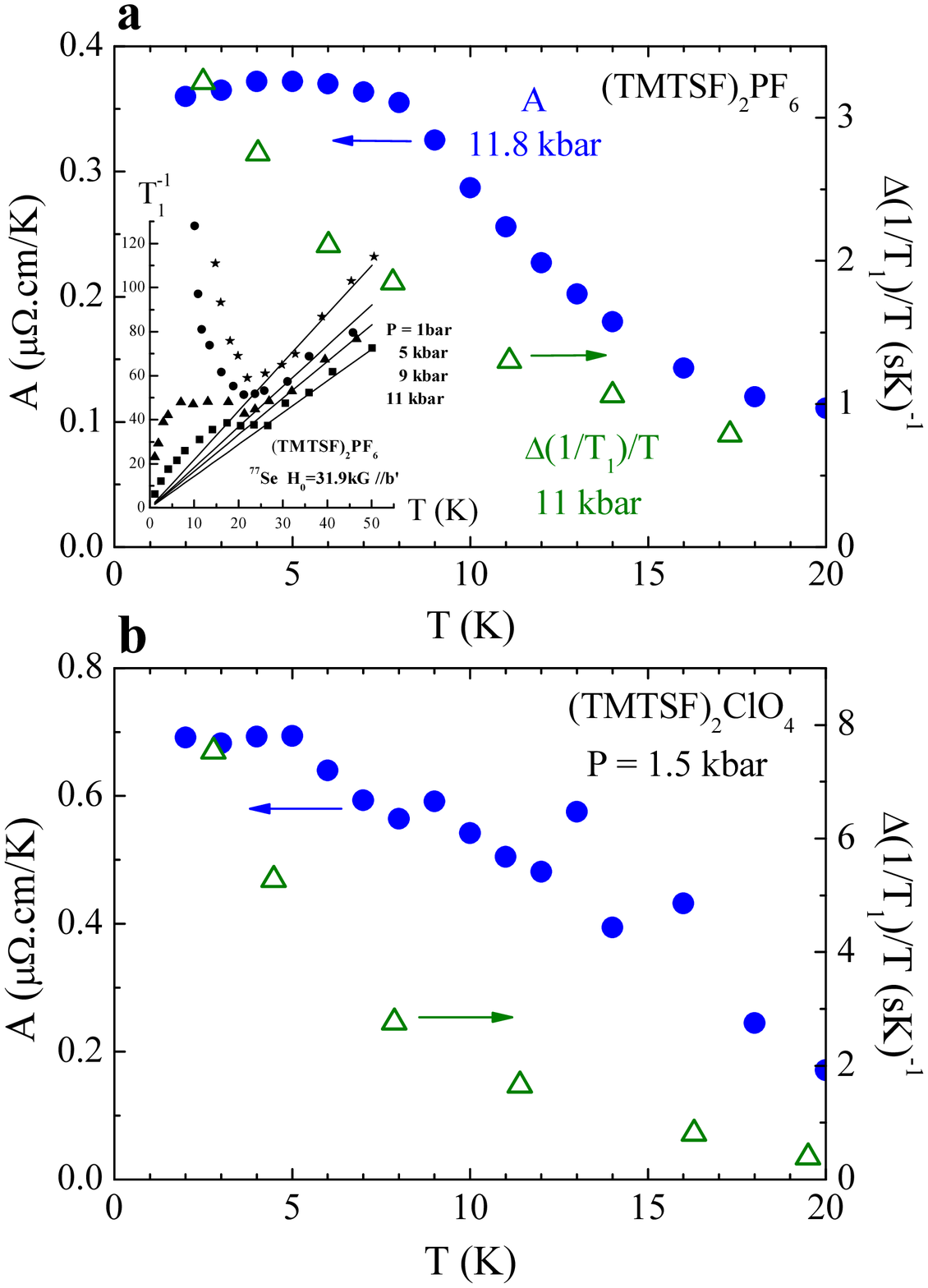}}
\caption{(a) Temperature dependence of the fluctuation-induced relaxation $\Delta(1/T_{1})/T$ at $11$ kbar deduced from NMR data in Ref.\cite{Bourbonnais88} and of the coefficient of the linear resistivity ($A$) at $11.8$ kbar for \tmp6. The insert shows the $^{77}$Se relaxation data at four different pressures where the bump of extra relaxation coming form AF fluctuations is clearly seen\cite{Bourbonnais88}. Similar results have been obtained in Ref.\cite{Brown08}. (b) Temperature dependence of the fluctuation-induced relaxation $\Delta (1/T_{1})/T$ at a pressure of $5$ kbar at room temperature, from Ref.\cite{Creuzet85a}, estimated to be around $1.5$ kbar at low temperature, and of the coefficient of the linear resistivity ($A$) at $1.5$ kbar for \tmc.
\label{Fig8_AT1TvsT}}
\end{figure}

The remarkable feature of the transport analyzed  according to the procedure presented above up to 20K is the temperature dependence of the prefactors in the temperature dependence of the resistivity when a second order polynomial form is assumed for its temperature dependence.
The linear term was shown to be related to the scattering of the carriers off AF fluctuations\cite{Doiron09} which are also active contributing to the nuclear spin-lattice relaxation adding a fluctuation contribution to the regular Korringa contribution \cite{Bourbonnais09} (see Sec.~\ref{Scaling}). A comparison between $A$ and the extra relaxation from AF spin fluctuations, subtracting the Korringa relaxation from the experimental data is shown on Fig.~\ref{Fig8_AT1TvsT} for \tmp6 at 11 kbar (a) and for \tmc around 1.5 kbar (b). 

However, the  comparison between  temperature dependencies of $A$ and $\Delta (1/T_{1}T)$ should not be taken at face value since transport and NMR experiments have  been conducted at  pressures of  11 and 11.8 kbar for NMR\cite{Bourbonnais88} and transport in \tmp6 respectively.  It is only a confirmation for  a common origin for the linear law in transport and the enhancement of relaxation due to the onset of AF fluctuations at low temperature.
For both compounds the singular carrier scattering and AF fluctuations go hand in hand in temperature and also under pressure.
 It is  tempting for \tmp6 to link the drop of the factor $B_{LT}$ (related to the regular  electron-electron Umklapp scattering),  which is observed close to \pc  (see Fig.\ref{Fig7_BChiP}-a)  to the opening of a pseudogap in the density of states related to fluctuations as observed in 1D conductors with a Peierls ground state\cite{Niedoba74}. If this were the case,  the amplitude for the residual resistivity coming from a fit of the high temperature quadratic regime with a constant plus a quadratic $T$ dependent contribution should be smaller (and not larger as observed experimentally) than the value for the residual resistivity given by the procedure presented  in this work.
 In addition, with the pseudogap scenario the residual resistivity should reveal a much stronger pressure dependence than what is actually observed  in the insert  of Fig.\ref{Fig4_DRavsTR0vsP}-a.

Moreover, as far as the spin sector is concerned,  no gap opens in the same temperature range according to   the Faraday susceptibility which is  nearly temperature independent  below 30 K in \tmc\cite{Miljak83,Miljak85} and also in NMR where relaxation data in \tmp6  do not reveal any precursor drop of the relaxation rate\cite{Wzietek93}.

\section{Theoretical aspects and connection  to experiment}
 \label{Scaling}
In  this section we  highlight some   results   of the renormalization group (RG)  method obtained   in the framework of the quasi-one-dimensional electron gas model. Their link with the properties of the Bechgaard salts and  their applicability  to the  resistivity data of the preceding sections will be  discussed.
\subsection{Previous results: phase diagram and nuclear relaxation}
The non interacting part of the quasi-one-dimensional electron gas model is defined in terms of a strongly anisotropic  electron spectrum yielding  an orthorhombic variant of the open Fermi surface in the $ab$ plane of   the Bechgaard salts.  The spectrum $E({\bf k}) = v_F(|k|-k_F) -2t_\perp\cos k_\perp - 2t_\perp'\cos 2k_\perp$  as a function of the momentum ${\bf k}=(k,k_\perp)$  is characterized by an intrachain or longitudinal Fermi energy $E_F=v_Fk_F$  which revolves around 3000~K in (TMTSF)$_2$X \cite{Ducasse86,Lepevelen01}; here  $v_F$ and $k_F$ are the longitudinal Fermi velocity and wave vector ($\hbar=1$ and $k_B=1$ throughout this section). This energy is much larger than the interchain hopping integral $t_\perp$ ($\approx 200$K),  in turn much bigger than the second-nearest neighbor transverse hopping amplitude $t_\perp'$. The latter stands as the antinesting parameter of the spectrum which simulates the main influence of pressure in the model.  The interchain hopping in the third and   less conducting direction is neglected. Electrons pertaining to different Fermi sheets are interacting through the backward and forward scattering amplitudes $g_1$ and $g_2$,  to which small longitudinal  Umklapp scattering  term  $g_3\ll g_1$, is added as a consequence of the slight dimerization  of the stacks and the anion potential \cite{Emery82}; here all interactions are normalized by the bandwidth $2E_F= \pi v_F$.

\begin{figure}[ht]
 \includegraphics[width=15.0cm]{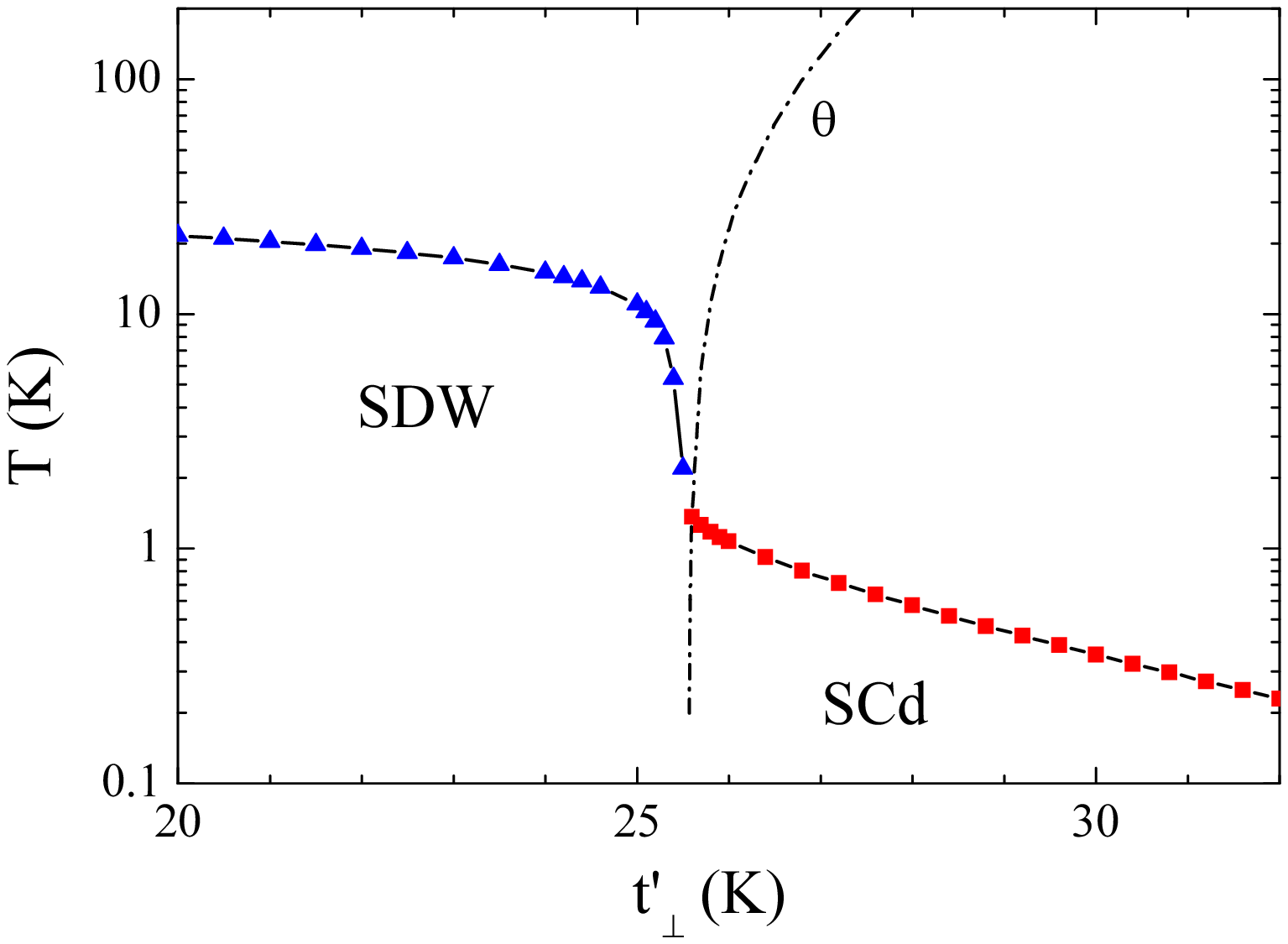}
 \caption{Calculated phase diagram of the quasi-one-dimensional electron gas model.  The scale $\Theta$ (dashed line) is the Curie-Weiss temperature.\label{Fig9_Dphases}} 
 \end{figure}

The interaction parameters that    shall be used in the following coincide with those previously employed in  the RG description of the phase diagram  and NMR spin-lattice relaxation rate \cite{Nickel06,Bourbonnais09}.   Taking $g_1=g_2/2 \approx 0.32$ and $g_3\approx 0.02$, the RG integration of high energy electronic degrees  of freedom is  carried out  down to the Fermi level, which leads to the flow or renormalization of the couplings at  the temperature $T$ \cite{Duprat01,Nickel06,Bourbonnais09}. At the one-loop level, the flow   superimposes  the $2k_F$ electron-hole (density-wave) and Cooper pairing many-body processes which combine and interfere    at every  order of perturbation. As a function of the `pressure' parameter  $t_\perp'$, a   singularity  in the scattering amplitudes signals an instability of the metallic state  toward the formation of an ordered state at some characteristic temperature scale. At low $t_\perp'$, nesting is sufficiently strong to induce an SDW instability in  the temperature range  of experimental  $T_{\rm SDW}\sim 10-20$~K. When the antinesting parameter approaches the threshold $t_\perp'^*$  from below  ($t_\perp'^* \approx 25.4~{\rm K}$, using the above  parameters), $T_{\rm SDW}$ sharply decreases and  as a result of interference, SDW correlations ensure  Cooper pairing attraction in the superconducting d-wave (SCd) channel. This   gives rise to  an    instability  of the normal state   for the onset of  SCd order at  the temperature $T_c$.  The  maximum   $T^*_c \approx 1.4$~K  is reached at $t_\perp'^*$, where    a steady decline  is initiated  as  $t_\perp'$ is further increased. The calculated scale for ordering  yields the phase diagram  of Fig.~\ref{Fig9_Dphases}, which reproduces   the  experimental trace for the onset  of long-range ordering in a system  like (TMTSF)$_2$PF$_6$ (Fig.~\ref{Fig1_DP3K}).
Another peculiar feature that comes out of the calculations  concerns the enhancement of spin correlations above $T_c$. It has been shown that despite strong alterations of  the nesting conditions and the existence of a singlet SCd ground state, the SDW susceptibility continues to grow as the temperature is lowered  following a Curie-Weiss law $\chi_{\rm SDW} \sim \xi^2$ where $\xi \propto (T+\Theta)^{-1/2}$ is the SDW correlation length\cite{Bourbonnais09}. This behavior reveals the   constructive feedback of Cooper  pairing on  antiferromagnetism, as mentioned above. The Curie-Weiss scale $\Theta\to 0$  is critical at $t_\perp'^*$, whereupon it undergoes a rapid  increase with $t_\perp'$ (Fig.~\ref{Fig9_Dphases}). Because of  the onset of superconductivity at   $T_c^*$, however,   the quantum critical behavior of  $\chi_{\rm SDW}\sim 1/T$ is avoided at  $t_\perp'^*$  \cite{Bourbonnais09}. 

The Curie-Weiss  behavior  has been shown to also govern the   enhancement of the nuclear relaxation rate $T_1^{-1}$ by SDW fluctuations. Deviations to the Korringa law  take  the form   $(T_1T)^{-1} \sim (T+\Theta)^{-1}$ at low temperature \cite{Bourbonnais09},  in accordance  with the experimental results  for(TMTSF)$_2$PF$_6 $ \cite{Wu05,Brown08,Creuzet87} (inset of Fig.~7) and (TMTSF)$_2$ClO$_4$ \cite{Shinagawa07,Creuzet85a,Bourbonnais84}.   The pressure dependence of $\Theta$ for both compounds is consistent with a critical reduction  of this scale as $P\to P_c$, namely  when  $T_c$  reaches its maximum value (Fig.~1-a).

\subsection{Quasi-particle scattering rate and resistivity\label{QP}}
The impact that can have   the correlations responsible of the structure of the phase diagram and the anomalous nuclear relaxation  on the single particle properties can be examined  by performing the RG procedure at the two-loop level. In Ref.~ \cite{Sedeki10}, one-particle self-energy  RG calculations  have been carried  out    using for the vertex part the contributions of the  mixed electron-electron and electron-hole scattering channels obtained in the one-loop level.  Thus the calculation  of the one-particle Matsubara self-energy,   $\Sigma_s({\bf k}_F(k_\perp),i\omega_n)$, allows us to extract the electron-electron scattering rate  at ${\bf k}_F(k_\perp)$  on the Fermi surface. The Fermi wave vector  ${\bf k}_F(k_\perp)$   is parametrized by the transverse wave vector   $k_\perp$. The scattering rate is obtained by the analytic continuation of   $\Sigma_s({\bf k}_F(k_\perp),i\omega_n)$     to the so-called  retarded  form $\Sigma_s({\bf k}_F(k_\perp),\omega+ i0^+)$ defined  on the real $\omega$ axis, and which consists of a real  ($\Sigma'_s$) and imaginary ($\Sigma''_s$) parts. From the zero frequency limit of the imaginary part,  one defined the  decay rate of quasi-particles on the Fermi surface $\tau^{-1}_{k_\perp}\equiv -2\Sigma_s''({\bf k}_F(k_\perp),\omega\to 0)$.

In the metallic state of the superconducting sector   $t_\perp'> t_\perp'^*$ of the phase diagram shown in Fig.~\ref{Fig9_Dphases}, the scattering rate turns out  to be anomalous in  both  $k_\perp$ and $T$ \cite{Sedeki10}. Marked deviations with respect to the Fermi liquid   behavior $\tau^{-1}_{k_\perp} \sim T^2 $ are found,  deviations whose amplitude strongly varies with $k_\perp$.  
  The absolute maximum is found in the longitudinal direction $k_\perp=0$, with secondary maxima    taking place at $k_\perp=\pm \pi$, namely  where the edges of the open Fermi surface cross  the  Brillouin zone in the perpendicular direction. These points markedly differ from the expected `hot'  spots at $k_\perp=\pm \pi/4$ and $\pm 3\pi/4$, as deduced from $E({\bf k})$ at   the best nesting conditions for the antiferromagnetic wave vector ${\bf q}_0=(2k_F,\pi)$. The anisotropy results    from  the interference of electron-hole with  electron-electron scattering,  which  moves the maxima in the regions where the superconducting SCd order parameter or the  gap  is expected to take  its largest values  below $T_c$.

Assuming that correlations over which electrons scatter are at equilibrium and can degrade momentum through Umklapp or impurity scattering,  the decay rate  will also affect conductivity. In the  relaxation time approximation, the contribution of the above singular  part of the self-energy  to the static conductivity at low temperature is given by
\begin{equation}
\sigma_s= {\omega_p^2\over 4\pi} \langle \tau_{k_\perp}\rangle,
\end{equation}
where $\langle \tau_{k_\perp}\rangle$ stands as the average of the relaxation time over the Fermi surface and $\omega_p$ is the longitudinal plasma frequency.
\begin{figure}[ht]
 \includegraphics[width=12.0cm]{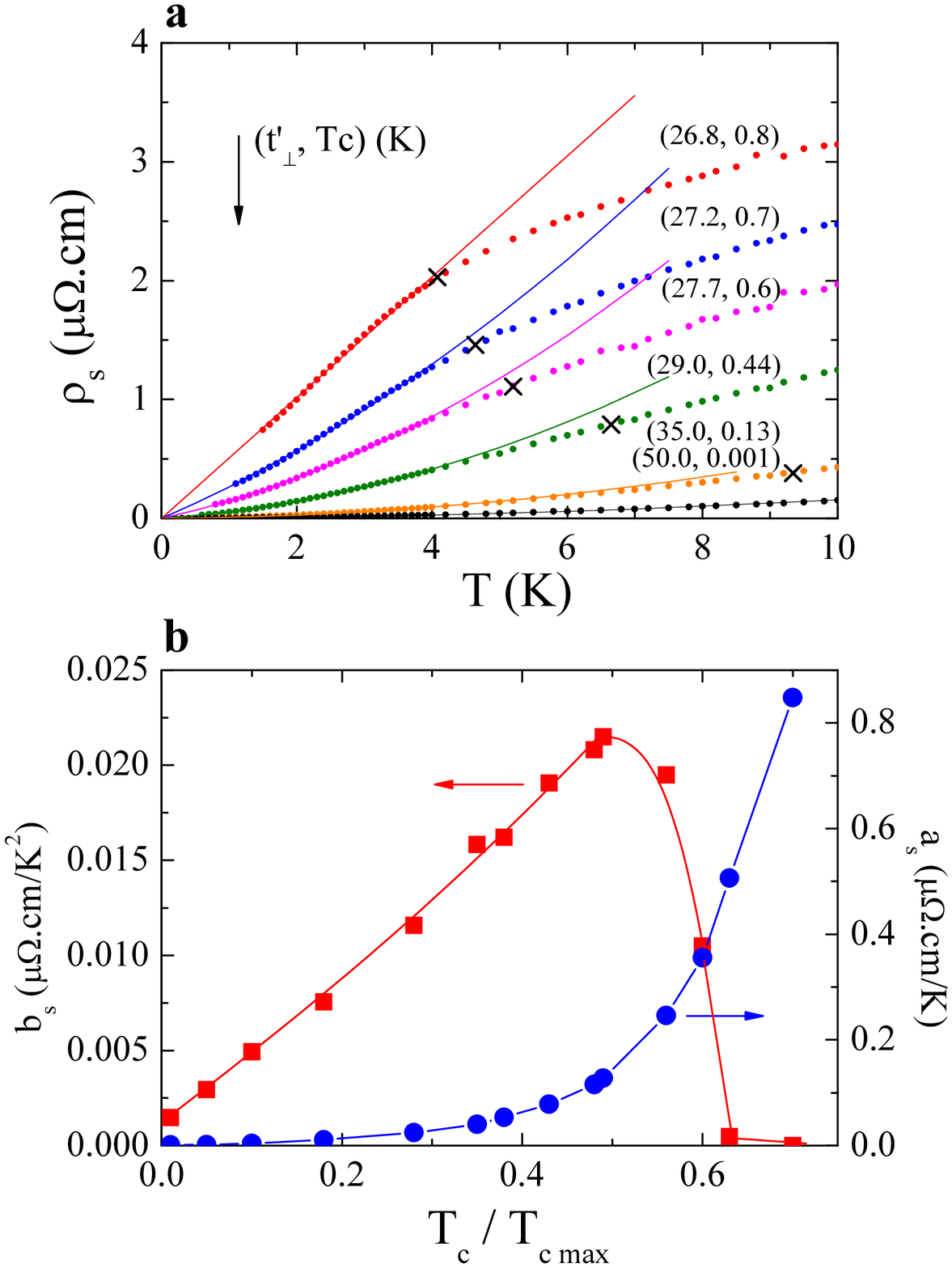}
 \caption{(a) Temperature dependence of  the calculated   resistivity from the singular part of the self-energy  for  the quasi-one-dimensional electron gas model in the superconducting sector. The continuous line is a fit to the expression (\ref{poly})  in the interval $4~{\rm K} >T>  1{\rm K}$ and  crosses refer to the scale $T_0$ below which the polynomial form prevails. (b) Variation of the linear  and quadratic coefficients $a_s$ and $b_s$ obtained from the fit of $\rho_s$ in (a) as a function of $T_c/T_{cmax}$. The lines between the points are guides to the eye. \label{RsTh}} 
 \end{figure}
 The expression of the  calculated resistivity  $\rho_s(T)=\sigma_s^{-1}(T)$, can be rewritten  as   $ \rho_s(T)= \rho_0  {\tau_0}/\langle \tau_{k_\perp}\rangle  $  using  the Drude formula for the residual resistivity $\rho_0= 4\pi/(\omega_p^2\tau_0)$ and the elastic scattering time $\tau_0$.  The value of $\rho_0$ (in $\mu \Omega {\rm cm}$) can be fixed  to the experimental data for  (TMTSF)$_2$PF$_6$ (Sec.~\ref{TrPF6}) and (TMTSF)$_2$ClO$_4$  (Sec.~\ref{TrClO4}), letting the normalization time scale  $\tau_0$ as a parameter  to be fixed. However, there  is a common  discrepancy as to the  actual value of $\tau_0$  in these materials\cite{Ng83,Dressel96,Jerome82}. Taking for example    $\rho_0 \sim  3~\mu \Omega {\rm cm}$, as a typical  range of residual  resistivity    above $P_c$  (Fig.~\ref{Fig3_RvsTABvsT}), and  the electron density of  $n\simeq 1.4 \times 10^{21}~{\rm cm}^{-3}$[\cite{Moser00}], one finds $\tau_0\approx 0.8 \times 10^{-12}$~sec using the Drude formula. On the other hand, the analysis of both the Drude peak in optical  conductivity near $T_c$ \cite{Ng83} and the  non-magnetic pair breaking effect on $T_c$ \cite{Joo05} yield    $\tau_0 \sim 10^{-11}$sec,  a significantly larger value.  In the following  we shall take  $\tau_0 = 2.5 \times 10^{-12}$ sec,  a  value that   falls within the above  bracket  and leads to an amplitude of calculated   resistivity  in  the range of observed values for $\rho_a -\rho_0$ above $T_c$ (Fig.~\ref{Fig3_RvsTABvsT}).
 \begin{figure}[ht]
 \includegraphics[width=12.0cm]{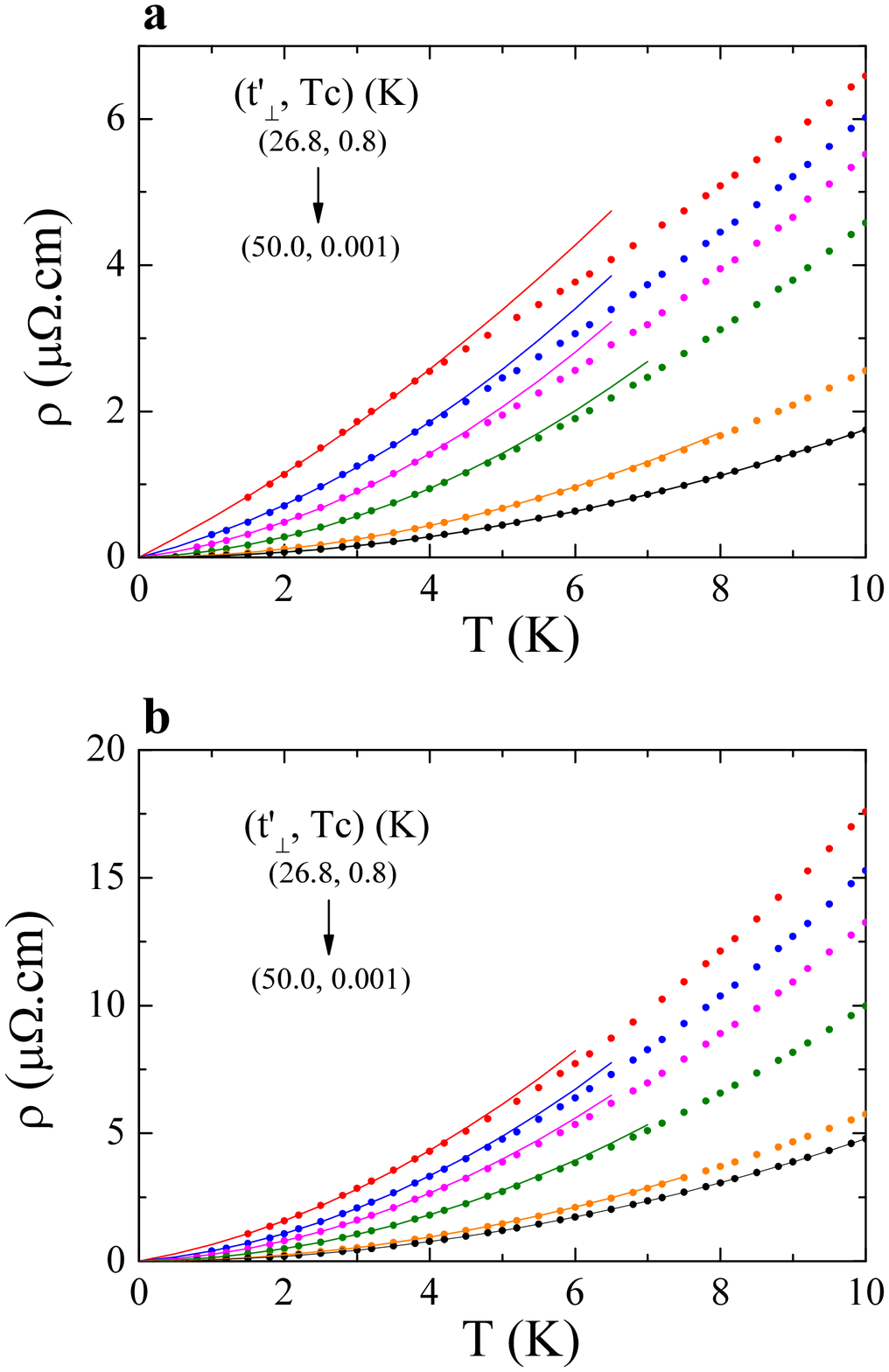}
 \caption{  Temperature dependence of  the calculated   resistivity including a residual Fermi liquid part  for the (TMTSF)$_2$PF$_6$ (a) and (TMTSF)$_2$ClO$_4$ (b) salts. The continuous lines correspond to a polynomial fit to $\rho(T) = aT + bT^2$  in the interval $4 >T>  1~{\rm K}$. \label{RTh}} 
\end{figure}
 
 The calculated temperature dependence of the parallel resistivity $\rho_s(T)=1/\sigma_s(T)$  is shown in Figure~\ref{RsTh}-a for different amplitudes of the  antinesting parameter  $t_\perp'$, which correspond to  different values of the  ratio  $T_c/T_{cmax}$.
 \begin{figure}[ht]
 \includegraphics[width=9.5cm]{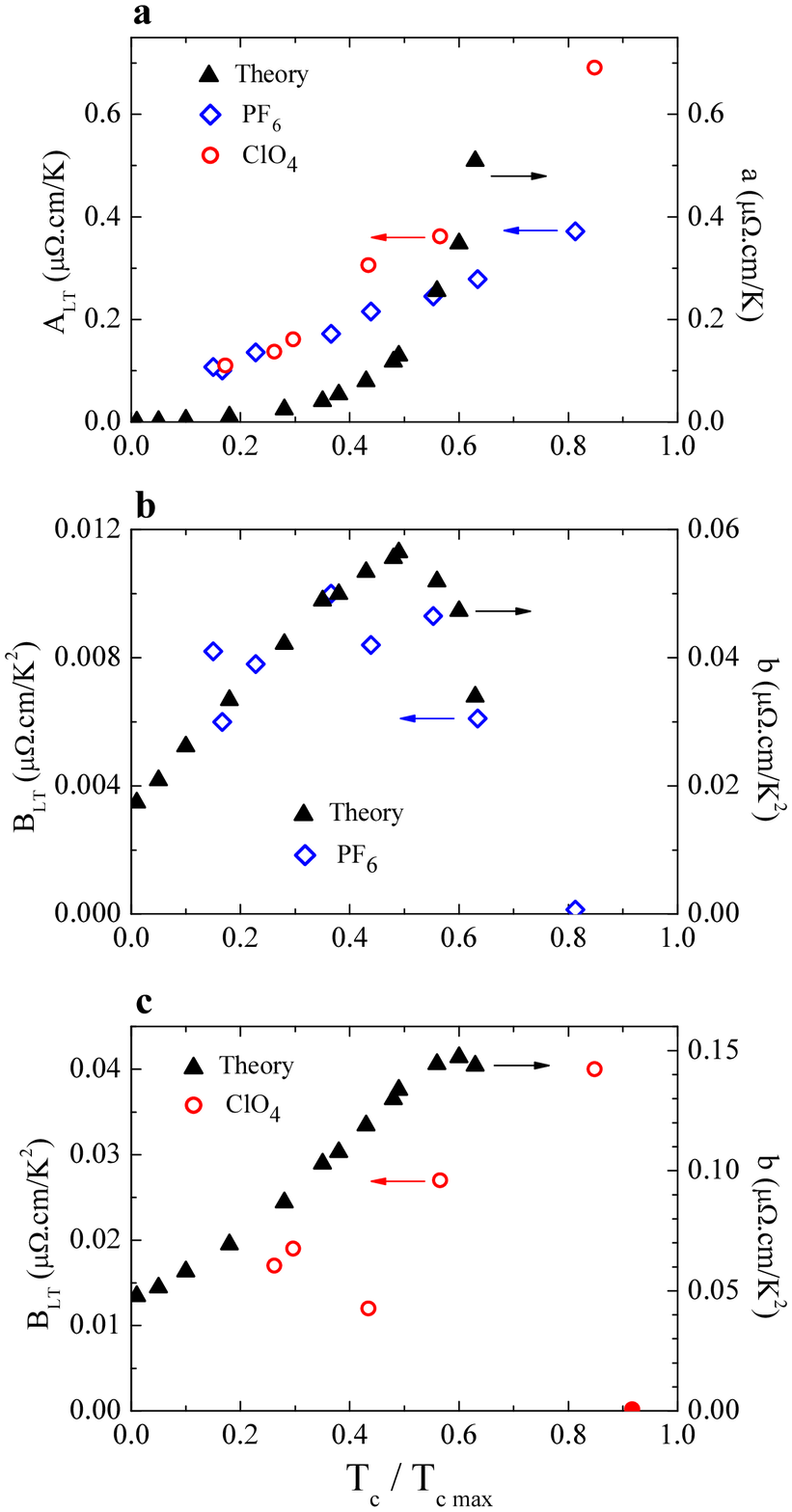}
 \caption{Variation of the calculated linear resistivity coefficient $a$, (a), and the Fermi liquid coefficient $b$ for (TMTSF)$_2$PF$_6$ (b) and (TMTSF)$_2$ClO$_4$ (c) as a function of the ratio $T_c/T_{cmax}$ (full triangles). The open diamonds (open circles) correspond to the experimental data of (TMTSF)$_2$PF$_6$ ((TMTSF)$_2$ClO$_4$). The experimental point at $T_c/T_{cmax}$=0.91 for \tmc refers to the ambient pressure data on Fig.\ref{Fig2_RTAO}.
\label{ABvsTc}} 
\end{figure}

Not shown in the Figure are the cases where $t_\perp' \sim t_\perp'^*$, namely close to the critical `pressure' in the phase diagram. In this domain  the fluctuations, for the most part antiferromagnetic,  become sufficiently pronounced to induce critical scattering and an insulating behavior at low temperature, despite the presence of a  superconducting ground state -- the paraconductive contribution to conductivity being excluded of the present calculations. While reminiscent of the reentrant region of  the actual phase diagram close to $P_c$ (Fig.~1-a), this behavior signals  the  flow  to strong coupling. 

 As $t_\perp'$  grows apart from  $t_\perp'^*$, the resistivity becomes metallic at low temperature.   It drops with an inward curvature down to the temperature scale $T_0$, below which it shows a downtrend toward zero. This  is displayed in Fig.~\ref{RsTh}-a down to the beginning of criticality close $T_c $ where the RG procedure is stopped.    Contrary to what is expected  for  a  Fermi liquid, however, the resistivity  is rather well described by the   polynomial form 
\begin{equation}
\label{poly}
 \rho_s(T)\approx a_sT + b_sT^2,
\end{equation}
  within the   interval $T_c\lesssim T< T_0$, which falls within  the Curie-Weiss  domain of spin correlations  discussed previously in the framework of NMR relaxation.   At the lowest  $t_\perp'$ shown in  Fig.~\ref{RsTh}-a  ($ 26.8~$K or $T_c\simeq 0.8~$K), the interval is     bound from below by  the onset of the superconducting critical fluctuations in the vicinity of   $T_c$ and   extends  up to $T_0 \approx 4$~K (cross in Fig.~\ref{RsTh}-a).  At that  $t_\perp'$, the temperature dependence  is essentially described by a linear term with  $b_s\approx0$ (Fig.~\ref{RsTh}-b). The existence  of   linearity  is the consequence of the scattering of electrons on  prominent antiferromagnetic fluctuations, though   strongly mixed with  Cooper SCd pairing. These fluctuations are peaked at    frequencies    smaller than   the temperature and as bosonic excitations, their coupling to electrons  in two dimensions is known to yield    a scattering rate $  \tau_{k_\perp}^{-1}\sim T\xi$, which  is  essentially  linear in temperature when the correlation length $\xi $  becomes temperature independent. This  is the case of the Curie-Weiss domain where $  \tau_{k_\perp}^{-1}\sim T $ at $T\ll \Theta$. The RG results presented here are compatible  with this limit.
  
   As $t_\perp'$ further increases, however, one gradually departs from this limit.  The temperature  domain where the resistivity follows the polynomial form  (\ref{poly}) expands    due to the growth of $T_0$ and the drop of $T_c$. While the  linear term  persists far from $t_\perp'^*$, the  $a_s$ coefficient steadily declines with the strength of  antiferromagnetic correlations  and $T_c$  (Fig.~\ref{RsTh}).
    The decrease  of $a_s\sim T_c^2$  approximately follows  the square of the critical temperature showing that both quantities are closely related (see Fig.~\ref{RsTh}-b). 
     A distinct  feature that comes out of the RG calculations of  Fig.~\ref{RsTh}-a    is the   emergence     of a $b_sT^2$ -- Fermi liquid -- term that accompanies the drop of $a_s$  within  the same temperature interval. The  $b_s$ factor first rises from zero,     reaches a maximum  to  finally decrease and level off toward  small values at  large $t_\perp'$  or vanishing  $T_c$ (Fig.~\ref{RsTh}-b). This interplay between the two terms is indicative of a progressive shift of the fluctuation spectral weight   to frequency scales  higher than  temperature. The stiffening of fluctuation frequency with respect to $T$ yields favorable conditions for establishing a Fermi liquid component \cite{Vilk97}.
     
     Interestingly,  the overall amplitude of resistivity  follows a trend similar to $T_c$ and vanishes at large $t_\perp'$, according to Fig.~\ref{RsTh}-a.  While this behavior adheres with the one found  in experiments for the strength of the inelastic contribution $\Delta \rho_a=\rho_a-\rho_0$  close to  $T_c$,  as shown in Figure~\ref{Fig3_RvsTABvsT} and in Ref.~\cite{Doiron09}, it differs  well  above this scale where  a sustained $T^2$ variation of resistivity is experimentally found.  It turns out, however,  that  the RG self-energy calculations considered so far  only  include contributions of the singular electron-hole and electron-electron scattering channels. The RG procedure actually neglects  all pieces of  residual   scattering that     in principle also yield a $T^2$ term for the quasi-particle decay rate and   resistivity --  a $T^2\ln T$ term  to be exact in strictly  two dimensions \cite{Hodges71,Gorkov98}.   The latter contribution eventually  takes over and dominates as the temperature is raised and  antiferromagnetic correlations  decrease enough in amplitude.  The situation is similar to the one found in the description of NMR experiments (e.g. inset of Figure~\ref{Fig8_AT1TvsT}, and Refs.\cite{Wzietek93, Bourbon89}), where non singular spin fluctuations, uniform to be specific,  give rise to  a Korringa like    temperature dependence  for the   nuclear spin relaxation rate ($(T_1T)^{-1} \propto \chi^2({\bf q}=0))$ Ref.~\cite{Bourbonnais09}.  As the temperature grows, this  Fermi liquid contribution   ultimately overcomes  the low-$T$ enhancement of $T_1^{-1}$ coming  from  antiferromagnetism.

The  regular contribution to the scattering rate can be in first approximation be considered decoupled from the singular part of the self-energy -- an approximation that has its limitations as we will see. It  can thus be added  to the expression (\ref{poly}), according to the usual  rule $\rho \to \rho_s(T) + b_rT^2$ ---  dropping the 2D  $\ln T$ correction, which has been assumed  to be cut  off by the   transverse hopping term in  the third direction.  We shall fix the coefficient $b_r$   by adjusting the calculated $\rho$ to the experimental values of  $\Delta \rho_a(15~{\rm K})$  obtained at 15~K  for (TMTSF)$_2$PF$_6$   and (TMTSF)$_2$ClO$_4$   at different pressures or $T_c$ (Figs.~\ref{Fig3_RvsTABvsT} and \ref{Fig4_DRavsTR0vsP}). The resulting trace of the   calculated resistivity  versus temperature is  shown in Fig.~\ref{RTh}  for different $t_\perp'$ (or $T_c$) in  the case of (TMTSF)$_2$PF$_6$ (Fig.~\ref{RTh}-a) and  (TMTSF)$_2$ClO$_4$ (Fig.~\ref{RTh}-b). The calculated data points have been   fitted to  the polynomial form $\rho(T) = aT + bT^2$ in the interval $  1\,{\rm K} < T <Ê 4\,{\rm K}$  (continuous curves in   Fig.~\ref{RsTh}-b), from which the variation of the  linear $a$ and $b$ coefficients with the ratio $T_c/T_{cmax}$ can be extracted  and compared to experiments.

The variation of the   linear   coefficient $a$ against    the ratio   $T_c/T_{cmax}$ is given in Fig.~\ref{ABvsTc}-a and compared with experimental results of Sec.~\ref{AContr} for the PF$_6$ and ClO$_4$ salts\cite{def}.  
It is clear here that the only contribution to the linear term comes from the singular part, so that $a\approx a_s$ and  essentially coincides with the one of Fig.~\ref{RsTh}-b. Thus compared to the values extracted   for both materials on experimental grounds, the calculated $a$ coefficient shows a  nearly quadratic variation  with $T_c$, which is   stronger than  found experimentally. The present theory, however, captures the progressive decay of the linear coefficient under `pressure'. The change is staggered over the entire domain of variation of $t_\perp'$   where $T_c$ differs noticeably from zero. The fact that a sizable amplitude of $a$ is not just confined to the very close proximity of $t_\perp'^*$ or $P_c$ on the pressure scale,   contrasts with what is  commonly expected near a quantum critical point \cite{Abanov03,Moriya90}. This  reveals that the  anomalous source of scattering for electrons   is only  gradually suppressed under `pressure',  leading  to  a broadened  interval of pressure over which quantum critical behavior can survive.  Extended criticality is  attributable  to the presence   of  Cooper pairing, which remains singular  and    interferes positively  with   antiferromagnetism; this provides  the link  between the linear  resistivity  and $T_c$. 

We now consider the  variation of the Fermi liquid coefficient $b$ as a function of $T_c/T_{cmax}$, which is  shown in Fig.~\ref{ABvsTc}-b for the PF$_6$ salt and in Fig.~\ref{ABvsTc}-c for ClO$_4$.  In the case of PF$_6$, this regular Fermi liquid behavior at high temperature is not strongly pronounced which lets the singular contribution of Fig.~\ref{RsTh}-b showing through the variation  of $b$ with $T_c$. A maximum of $b$ at intermediate $T_c$  is thus found,   an important result  that is manifest in (TMTSF)$_2$PF$_6$. 
In the case of (TMTSF)$_2$ClO$_4$, the Fermi liquid term at high temperature is about three times stronger, leading to a larger regular $b_rT^2$ contribution. This  removes some weight  of the Fermi liquid $b_sT^2$ component coming from the singular term, and as a result, the maximum in $b$ is scarcely seen,  being masked for the most part by the contribution of the regular contribution.

\section{Conclusion}

In summary, this work reports a careful investigation of the metallic phase of   \tmp6 and \tmc carried out  under pressure which confirms the role played by high pressure controlling the size of electron correlations and in turn the stability of superconductivity in these compounds. However, this new study is pointing out a salient difference between the diagrams of these two superconductors. An  important control parameter of superconductivity in \tmtsfx  is also the value of the elastic electron life time. Such a lifetime is governed in \tmc  by the level of residual non magnetic impurities  originating  from the  imperfect anion ordering under high pressure conditions.
The possibility to achieve a large number of high pressure runs using a single  sample has enabled an exhaustive and quantitative
analysis  of longitudinal transport experiments  in the temperature regime 0-20 K in a wide pressure regime. 
Processing  the transport data using a sliding fit procedure in temperature according to  a second order polynomial form, $\rho(T) = \rho_0 + AT + BT^2$  and $T$ dependent prefactors, reveals two temperature domains: a high temperature domain ($T\approx$ 20 K) in which the regular $T^2$ electron-electron Umklapp scattering obeys a Kadowaki-Woods law and a low temperature regime ($T< 8$K) in which the scattering of carriers against antiferromagnetic fluctuations  provides for \tmp6 a purely linear $T$ dependent contribution\cite{Doiron09}. 

This  linear in temperature component of transport is connected to the intrinsic pressure dependence of \tc which is controlled by pressure while an  additional control parameter  of \tc  is given by the residual anion disorder in the case of \tmc. 
In both compounds a correlated behavior  exists between the linear term of transport and the extra nuclear spin-lattice relaxation due to antiferromagnetic fluctuations. 

In \tmp6, where the temperature dependence of the resistivity is likely cleaner than for \tmc  since it is free from the pollution of the \cl \, anions  ordering over a wide $T$ regime, a \textit{purely} linear in $T$ dependence ($B \rightarrow  0$) extending up to 8 K supports the  vanishing of the regular part of the quadratic scattering but does not imply the opening of a pseudo gap in the charge sector.
 The theoretical treatment of fluctuations in the case of channel mixing will require  additional work.

Other studies of  metallic and superconducting phases of these  compounds would be highly desirable in the future, in particular a reinvestigation of the far-infrared properties, additional NMR work under pressure in both compounds and a detailed investigation of transport in the vicinity of the critical pressure in \tmp6.

We have compared in some details the experimental data to the predictions of the two-loop scaling theory for the resistivity, as it can be derived from the quasi-particle scattering rate within the Boltzmann picture.  A  low temperature  linear  term for resistivity emerges naturally from the theory; its  amplitude  peaks near the critical point  and is followed by a gradual decay  extending  over  the entire range of pressure where a -- d-wave -- superconducting $T_c$ differs from zero. This  remarkable  feature  co-occurs   with another one, namely the low temperature development of Fermi liquid scattering   whose pressure dependence is  for the most part opposed to the one of the linear component of resistivity. Both results are  finding a  favorable echo in a material like (TMTSF)$_2$PF$_6$,  and to a large extent  in  (TMTSF)$_2$ClO$_4$ as well.

The correlation between linear resistivity and $T_c$ can definitely  be regarded as  another important result of the theory. It  supplies some  basic microscopic insight on  the behavior of electron scattering in the presence of  antiferromagnetic fluctuations and Cooper pairing, a  long-established problem in unconventional superconductors. In the  quasi-one dimensional electron gas  these two channels of correlations are intrinsically  interdependent, which is systematically  taken into account within scaling theory.  This compound pairing turns out to be of crucial importance in matching the pressure  range of linear resistivity to the one of $T_c$. The view developed in this work about electrical resistivity proves to  be internally consistent  with the previous analysis made in the context of the nuclear relaxation and the phase diagram using the same  approach.  This    gives  important additional support to the model proposed.

This work was supported by NSERC (Canada), FQRNT (Qu\'ebec), CFI (Canada), a Canada Research Chair (L.T.),  the Canadian Institute for Advanced Research and CNRS (France).


\begin{thebibliography}{10}
\bibitem{def}
$T_{cmax}$ is the maximum value of $T_{c}$ observed for a \tmc sample coming
  from the same batch with a measurement along the $c^{\star}$
  direction\cite{Yonezawa08}. In case of \tmp6, $T_{cmax}$ is the value for
  $T_{c}$ in the same sample measured under 8.4 kbar in the coexistence domain.
\end{thebibliography}
\end{document}